\def\real{{\tt I\kern-.2em{R}}}  
\def\nat{{\tt I\kern-.2em{N}}}     
\def\eps{\epsilon} 


\def\realp#1{{\tt I\kern-.2em{R}}^#1}
\def\natp#1{{\tt I\kern-.2em{N}}^#1}
\def\hyper#1{\,^*\kern-.2em{#1}}
\def\monad#1{\mu (#1)}

\def\St#1{{\tt st}#1}
\def\st#1{{\tt st}(#1)}
\def\hyperreal{{^*{\real}}}
\def\hyperrealp#1{{\tt ^*{I\kern-.2em{R}}}^#1} 

\def\hypernatp#1{{{^*{{\tt I\kern-.2em{N}}}}}^#1} 
\def\eskip{\hskip.25em\relax}

\def\Hyper#1{\hyper {\eskip #1}}
\def\leaderfill{\leaders\hbox to 1em{\hss.\hss}\hfill}
\def\srealp#1{{\rm I\kern-.2em{R}}^#1}

\def\pars{\par\smallskip}
\def\parm{\par\medskip}

\def\b#1{{\bf #1}}
\def\ref#1{$^{#1}$}

\def\m@th{\mathsurround=0pt}
\def\rightarrowfill{$\m@th \mathord- \mkern-6mu \cleaders\hbox{$\mkern-2mu 
\mathord- \mkern-2mu$}\hfil \mkern-6mu \mathord\rightarrow$}
\def\leftarrowfill{$\mathord\leftarrow
\mkern -6mu \m@th \mathord- \mkern-6mu \cleaders\hbox{$\mkern-2mu 
\mathord- \mkern-2mu$}\hfil $}
\def\noarrowfill{$\m@th \mathord- \mkern-6mu \cleaders\hbox{$\mkern-2mu 
\mathord- \mkern-2mu$}\hfil$}
\def\orgate{$\bigcirc \kern-.80em \lor$}
\def\andgate{$\bigcirc \kern-.80em \land$}
\def\inverter{$\bigcirc \kern-.80em \neg$}
 \magnification=1000
\tolerance 10000
\hoffset=0.25in
\hsize 6.00 true in
\vsize 8.75 true in

\pageno=1

\baselineskip 13pt
\font\eightrm=cmr9
\centerline{{\bf A Nonstandard Derivation for the Special Theory of Relativity}\footnote*{This is an expanded version of the paper presented at the 14 NOV 1992 Meeting of the Mathematical Association of America, Coppin State College, Baltimore MD and as it appears in  Herrmann$,$ R. A.$,$ Special relativity and a nonstandard substratum$,$ {\it Speculat. Sci. Technol.$,$} {\bf 17}(1994)$,$ 2-10.}} \bigskip
\centerline{Robert A. Herrmann}\parm
\centerline{Mathematics Department}
\centerline{U. S. Naval Academy}
\centerline{572C Holloway Rd}
\centerline{Annapolis$,$ MD 21402-5002 USA}
\centerline{21 SEP 1993}\bigskip\par
{\leftskip = 0.5in \rightskip = 0.5in \noindent {\eightrm {\it Abstract:}} Using properties of the 
nonstandard physical world$,$ a new fundamental derivation for  
effects of the Special Theory of Relativity is given. This fundamental 
derivation removes all the contradictions and logical errors in the original 
derivation and leads to the fundamental expressions for the Special Theory Lorentz transformation. Necessarily, these are obtained by means of hyperbolic geometry.  
It is shown that the Special Theory effects are manifestations of the 
interaction between our natural world and a nonstandard medium, the NSPPM.  This 
derivation  eliminates the controversy associated with any physically  
unexplained absolute time dilation and length contraction. It is shown that 
there is no such thing as a absolute time dilation and length contraction 
but$,$ rather$,$ alterations in pure numerical quantities associated with an 
electromagnetic interaction with an NSP-world NSPPM.\par}\par
\medskip
\leftline{\bf 1. The Fundamental Postulates.} 
\medskip
There are various Principles of Relativity. The most general and least 
justified is the one as stated by Dingle ``{\it There is no meaning in 
absolute motion.} By saying that such motion  has {\it no meaning}$,$
 we assert that 
there is no observable effect by which we can determine whether an object is 
absolutely at rest or in motion$,$ or whether it is moving with one velocity or 
another.''[1:1] Then we have Einstein's statements that  ``I. The laws of 
motion are equally valid for all inertial frames of reference. II. The velocity 
of light is invariant for all inertial systems$,$ being independent of the 
velocity of its source; more exactly$,$ the measure of this velocity (of light) 
is constant$,$ $c$$,$ for all observers.''[7:6--7] I point out that Einstein's 
original derivation in his 1905 paper ({\it Ann. der Phys.} {\bf 17}: 
891) uses certain well-known processes related to partial differential 
calculus.\pars 
In 1981 [5] and 1991 [2]$,$  
it was discovered that 
the intuitive concepts associated with the  Newtonian laws of motion were 
inconsistent with respect to the mathematical theory of infinitesimals when 
applied to a theory for light propagation. The apparent nonballistic nature for light propagation when transferred to infinitesimal world would 
also yield a nonballistic  behavior. 
Consequently$,$ {\bf there is an absolute contradiction between Einstein's 
postulate II and the derivation employed.} This contradiction would not have 
occurred if it had not been assumed that the \ae{ther} followed the 
principles of Newtonian physics with respect to electromagnetic propagation. 
[Note: On Nov. 14$,$ 1992$,$ when the information in this article was formally presented$,$ I listed various predicates that Einstein used and showed the specific places within the derivations where the predicate's domain was altered without any additional argument. Thus$,$ I gave specific examples of  
the model theoretic error of 
generalization.] 
\pars
 I mention that Lorentz speculated that \ae{ther} 
theory need not correspond directly to the mathematical structure but could 
not show what the correct correspondence would be. Indeed$,$ if one assumes that the 
{\bf Nonstandard Photon-particle Medium}, the NSPPM, satisfies the most basic concept associated with an 
inertial system that a body can be considered in a state of rest or 
uniform motion 
unless acted upon by a force$,$ then the expression $F=ma,$ among others$,$ 
may be altered for infinitesimal NS-substratum behavior. Further$,$ the 
NS-substratum$,$ when light propagation is discussed$,$ does not follow the
Galilean rules for velocity composition. The additive rules are followed but 
no negative real velocities exterior to 
the Euclidean monads are used since we are only interested in the propagation properties for electromagnetic radiation. 
The derivation in 
section 3 removes all contradictions by applying the most simplistic Galilean 
properties of motion$,$ including the ballistic property$,$ but only to behavior 
within a Euclidean monad. \pars
As discussed in section 3$,$ the use of an NSP-world (i.e. {\it nonstandard physical world}) 
NSPPM allows for the  elimination of the well-known 
Special Theory ``interpretation'' contradiction that the 
mathematical model 
uses the concepts
of Newtonian absolute time and space$,$ and$,$
yet$,$ one of the major interpretations is 
that there is no such thing as absolute time or absolute space.\pars 
Certain  general principles for NSPPM light propagation will be  
specifically stated  
in section 3.  These principles  
can be gathered together as follows: (1) {\sl There is a portion of the 
nonstandard photon-particle medium - the NSPPM - that sustains 
N-world  (i.e. natural = physical world) electromagnetic propagation. Such propagation 
follows the infinitesimally presented laws of Galilean dynamics$,$ when 
restricted to monadic clusters$,$  
and the 
monadic clusters follow an additive and an actual metric property 
for linear relative motion when considered collectively.} [The term 
``nonstandard electromagnetic field'' should only be construed as a NSPPM notion, where the propagation of electromagnetic radiation follows 
slightly different principles than within the natural world.] 
(2) {\sl The motion of light-clocks
 within the N-world (natural world) is associated with one single 
effect. This effect 
is an alteration in an appropriate light-clock mechanism.}  [The 
light-clock concept will be explicitly defined at the end of section 3.] 
It will be shown 
later that an actual physical cause may be associated with vreified Special Theory physical alterations. 
 {Thus the 
Principle 
of Relativity$,$ in its general form$,$ and the inconsistent
portions of the Einstein principles are
eliminated from consideration and$,$ as will be shown$,$ 
the existence of a special type of medium can be assumed without 
contradicting  experimental evidence.} \pars
In modern Special Theory interpretations [6]$,$ it is  claimed 
that the effect of ``length contraction''  has no physical meaning$,$ whereas 
time dilation does. 
This is probably true if$,$ indeed$,$ the Special Theory is actually based upon 
the intrinsic N-world  concepts of length and time. What follows will further
demonstrate that 
 the  Special Theory is  a light propagation theory$,$ 
as has been previously 
argued by others$,$ and that the so-called ``length contraction'' and time 
dilation can both be 
interpreted as  physically real effects when they are described in terms of the NSPPM. The effects are only relative to a theory of light 
propagation.\pars   
\medskip
\leftline{\bf2. Pre-derivation Comments.} 
\medskip
Recently [2]--[4]$,$ nonstandard analysis [8] has proved to be 
a very significant tool in investigating the mathematical foundations for various 
physical theories. In 1988 [4]$,$ we discussed how the methods of 
nonstandard analysis$,$ when applied to the symbols that appear in statements 
from a physical theory$,$ lead formally to a pregeometry and the entities termed 
as subparticles. One of the goals of  
 NSP-world research is the 
re-examination of the foundations for various controversial 
N-world theories and the eventual elimination of such 
controversies by viewing such theories as but restrictions of more simplistic 
NSP-world concepts. This also leads to indirect evidence for the actual 
existence of the NSP-world. \par
The Special Theory of Relativity still remains a very controversial theory due 
to its philosophical implications. Prokhovnik [7] produced a derivation that 
yields all of the appropriate transformation formulas based upon a light 
propagation theory$,$ but unnecessarily includes an interpretation of 
the so-called Hubble textural expansion  of our 
universe as an additional ingredient. 
The new derivation we give in this article shows that  
properties of a NSPPM also lead to 
Prokhovnik's expression  
(6.3.2) in reference [7] and from which all of the appropriate equations 
can 
be derived. However$,$ rather than considering the Hubble expansion as directly 
related to Special Relativity$,$ it is shown that one only needs to consider 
simplistic 
NSP-world behavior for light propagation and the measurement of 
time by means of N-world light-clocks. This leads to the 
conclusion that Special Theory effects may be produced by a dense NSPPM  
within the  NSP-world. Such an NSPPM -- an \ae{ther} -- 
yields
N-world 
Special Theory effects. \pars   
\medskip
\leftline{\bf 3. The derivation}
\medskip
The major natural system in which we exist locally is a space-time 
system. {``Empty'' space-time} has only a few characterizations when viewed 
from an Euclidean perspective. We investigate$,$ from the NSP-world viewpoint$,$ 
 {electromagnetic propagation through a Euclidean neighborhood} of space-time. 
Further$,$ we assume that light is such a propagation. One of the basic precepts 
of infinitesimal modeling is the experimentally verified {{\it simplicity}} 
for such a local system. For actual time intervals$,$ certain physical processes 
take on simplistic descriptions. These NSP-world descriptions are represented 
by the exact same description restricted to infinitesimal intervals. Let 
$[a,b],\ a \not= b,\ a > 0,$ be an objectively real conceptional time interval and let $t 
\in (a,b).$\par
The term  ``time'' as used above is very misunderstood. There are 
various viewpoints relative to its use within mathematics. Often$,$ it is but 
a term used in mathematical modeling$,$ especially within the calculus. It is a 
catalyst so to speak. It is a modeling technique used due to the necessity for 
infinitesimalizing physical measures. The idealized concept for the 
``smoothed out'' model for distance measure appears acceptable. Such an 
acceptance comes from the  use of the calculus in such areas as quantum 
electrodynamics where it has great predictive power. In the subatomic region$,$
the assumption that geometric measures have physical meaning$,$ even without 
the ability to measure by external means, is justified as an appropriate 
modeling technique. Mathematical procedures applied to 
regions ``smaller than'' those dictated by the uncertainty principle are 
accepted although the reality of the infinitesimals themselves need not be 
assumed. On the other hand$,$ for this modeling technique to be applied$,$ the 
rules for ideal infinitesimalizing should be followed.\par
 The infinitesimalizing 
of ideal geometric measures is allowed. But$,$ with respect to the time concept 
this is not the case. Defining measurements of time as represented by the 
measurements of some physical periodic process is not the definition upon 
which the calculus is built. Indeed$,$ such processes cannot be 
infinitesimalized. To infinitesimalize a physical measurement using physical 
entities$,$ the entities being  
observed must be capable of being smoothed out in an ideal sense. 
This means that only the macroscopic is considered$,$ the atomic or microscopic 
is ignored. Under this condition$,$ you must be able to subdivide the device  
into ``smaller and smaller'' pieces. The behavior of these pieces can then be 
transferred to the world of the infinitesimals. Newton based the calculus not 
upon geometric abstractions but upon observable mechanical behavior. It was 
this mechanical behavior that Newton used to define physical quantities that 
could be infinitesimalized. This includes the definition of ``time.''\par
All of Newton's ideas are based upon velocities as the defining concept. 
The notation that uniform (constant) velocity exists for an object when that object is 
not affected by anything$,$ is the foundation for his mechanical observations.
This is an ideal velocity$,$ a universal velocity concept. The modern approach 
would be to add the term ``measured'' to this mechanical concept. This will 
not change the concept$,$ but it will make it more relative to natural world 
processes and a required theory of measure. This velocity concept is coupled 
with a smoothed out scale$,$ a ruler$,$ for measurement of distance. Such a 
ruler can be infinitesimalized. From observation$,$ Newton then 
infinitesimalized his uniform velocity concept. This produces the theory of 
fluxions. \par
Where does observer time come into this picture? 
It is simply a defined quantity based upon the length and velocity concept. 
Observationally$,$ it is the ``thing'' we call time that has passed when a test 
particle with uniform velocity first crosses a point marked on a scale and 
then crosses a second point marked on the same scale. This is in the absence of any 
physical process that will alter either the constant velocity or the scale.
Again this definition would need to be refined by inserting the word 
``measured.'' Absolute time is the concept that is being measured and cannot be altered as aconcept. \par
Now with Einstein relativity$,$ we are told that measured quantities are effected 
by various physical processes. All theories must be operational in that the 
concept of measure must be included. But$,$ the calculus is used. Indeed, used by Einstein in his original derivation. Thus$,$ 
unless there is an actual physical entity that can be substituted for the 
Newton's ideal velocity$,$ then any infinitesimalizing process would 
contradict the actual rules of application of the calculus to the most basic 
of physical measures. But$,$ the calculus is used to calculate the measured 
quantities. Hence$,$ we are in a quandary. Either there is no physical basis for 
mathematical models based upon the calculus$,$ and hence only selected 
portions can be realized while other selected portions are simply parameters 
not related to reality in any manner$,$ or the calculus is the incorrect 
mathematical structure for the calculations. Fortunately$,$ nature has provided 
us with the answer as to why the calculus$,$ when properly interpreted$,$ remains 
such a powerful tool to calculate the measures that describe observed physical 
behavior.\par   
In the 1930s$,$ it was realized that the measured uniform velocity of the 
to-and-fro velocity of electromagnetic radiation$,$ (i.e. light) is the only 
known natural entity that will satisfy the Newtonian requirements for an ideal 
velocity and the concepts of space-time and from which the concept of time 
itself can be defined.  The first to utilize this in relativity theory was 
Milne. This fact I learned after the first draughts of this paper were
written and gives historical 
verification of this paper's conclusions. Although$,$ it might be assumed that 
such a uniform velocity concept as the velocity of light or light paths {\it in vacuo} 
cannot be infinitesimalized$,$ this is not the case. Such infinitesimalizing 
occurs for light-clocks and from the simple process of ``scale changing'' for a smoothed out ruler. What this means is that$,$ at its most basic {\it physical} level$,$ {\it conceptually} absolute or universal Newton time can have operational meaning as a physical foundation for a restricted form of ``time'' that can be used within the calculus. \par

As H. Dingle states it$,$ ``The second point is that the conformability of light to 
Newton mechanics . . . makes it possible to define corresponding units of 
space and time in terms of light instead of Newton's hypothetical `uniformly 
moving body.' '' [The Relativity of Time$,$ {\it Nature$,$} 144(1939): 888--890.]
It was Milne who first (1933) attempted$,$ for the Special Theory$,$ to use this definition for a ``Kinematic 
Relativity'' [{\it Kinematic Relativity}$,$ Oxford University Press$,$ Oxford$,$ 
1948] but failed to extend it successfully to the space-time environment. 
In what follows such an operational 
 time concept is being used and infinitesimalized. 
It will be seen$,$ however$,$ that based 
upon this absolute time concept another time notion is defined$,$ and this is the actual 
time notion that must be used to account for the physical changes that 
seem to occur due to relativistic processes. In practice$,$ the absolute time  is 
eliminated from the calculations and is replaced by defined ``Einstein time.'' 
It is shown that Einstein time can be infinitesimalized through the use of the 
definable  ``infinitesimal light-clocks'' and gives an exact measurement.\par

 Our first assumption is based entirely upon the logic of 
infinitesimal analysis$,$ reasoning$,$ modeling and subparticle theory. \parm 
{\leftskip 0.5in \rightskip 0.5in \noindent (i) { ``Empty'' space within 
our universe$,$ from the NSP-world viewpoint$,$ is composed of a dense-like 
nonstandard medium (the NSPPM) that sustains$,$ comprises and yields  N-world Special Theory  
effects. These NSPPM effects are electromagnetic in character.} \par}\parm 
\noindent This medium through which 
the effects appear to propagate comprise the objects that yield 
these effects. The next assumption is convincingly obtained from a simple and 
literal translation of the concept of infinitesimal reasoning. \parm
 {\leftskip 0.5in \rightskip 0.5in \noindent  (ii) {Any 
N-world position from or through which an electromagnetic effect appears to 
propagate$,$ when viewed from the NSP-world$,$ is embedded into a disjoint 
``monadic cluster'' of the NSMP, where this monadic cluster 
mirrors the same unusual order properties$,$ with respect to propagation$,$ as the 
nonstandard ordering of the nonarchimedian field of hyperreal numbers 
$\hyperreal.$ [2] A monadic cluster may be 
a set of NS-substratum subparticles located within a monad of the standard 
N-world position. The propagation properties within each such monad are 
identical.}\par}\parm 
In what follows$,$ consider two (local) fundamental pairs of N-world 
positions $F_1,\ F_2$ 
that are in nonzero uniform (constant) NSP-world linear and 
relative motion. Our interest is in what effect such nonzero velocity might 
have upon such electromagnetic propagation. Within the NSP-world$,$ this uniform and linear 
motion is measured by the number $w$ that is near to a standard number 
$\omega$ and this velocity is measured with respect to conceptional NSP-world time 
and a stationary subparticle field. [Note that field expansion can be additionally incorporated.] The same NSP-world linear ruler is used in 
both the NSP-world and the N-world. The only difference is that the ruler is 
restricted to the N-world when such measurements are made. 
N-world  time is  
measured by only one type of machine -- the light-clock. The concept of the 
light-clock is to be considered as any clock-like apparatus that utilizes 
either  
directly or indirectly an equivalent process. As it will be 
detailed$,$ due to the   
different propagation effects of electromagnetic radiation within the 
two  ``worlds$,$'' measured N-world light-clock time need not be the same 
as the NSP-world 
time. Further$,$ the NSP-world ruler is the measure used to define the 
N-world light-clock. \pars
Experiments show that for small time intervals $[a,b]$  
the {Galilean theory} of average velocities (velocitys) suffices to give 
accurate information relative to the compositions of such velocities. Let there 
be an internal function $q \colon \Hyper [a,b] \to \hyperreal,$ where  $q$ 
represents in the NSP-world a {distance function}. Also$,$ let nonnegative  and 
internal $\ell\colon \Hyper [a,b] \to \hyperreal$  be a function 
that yields the NSP-world 
velocity of the electromagnetic propagation at any  $t \in \Hyper [a,b].$ As usual 
$\monad t$ denotes the monad of standard $t,$ where ``$t$'' is an absolute NSP-world ``time'' parameter. \par

The general and correct methods of infinitesimal modeling state that$,$ within the internal 
portion of the NSP-world$,$ two measures $m_1$ and $m_2$ are {{\it 
indistinguishable for dt}} (i.e. infinitely close of order one) (notation $m_1 \sim m_2$) if and only if 
$0\not= dt \in \monad 0,$ ($\monad 0$ the set of infinitesimals)
$${{m_1}\over {dt}} - {{m_2}\over{dt}} \in \monad 0. \eqno(3.1)$$
Intuitively$,$ indistinguishable in this sense means that$,$ although within the 
NSP-world the two measures are only equivalent and not necessarily equal$,$ 
the {{\it first 
level}} (or first-order) effects  these measures represent over $dt$ are 
indistinguishable within 
the N-world (i.e. they appear to be equal.) \par
In the following discussion, we continue to use {{\it photon}}
 terminology. Within the N-world our photons need not be conceived of 
as particles in the sense that there is a nonzero finite N-world distance 
between individual photons. Our photons {\it may be} finite combinations of 
intermediate subparticles that exhibit$,$ when the standard part operator is 
applied$,$ basic electromagnetic field properties. They need not be discrete 
objects when viewed from the N-world$,$ but rather they could just as well give 
the {\it appearance} of a dense NS-substratum. Of course$,$ this dense NSPPM portion  
is not the usual notion of an  ``\ae{ther}'' (i.e. ether) 
for it is not a subset of 
the N-world. This dense-like portion of the NS-substratum contains{{\it nonstandard 
particle medium} (NSPPM)}. Again ``photon'' can be considered as 
but a convenient term used to discuss electromagnetic propagation. 
Now for another of our simplistic physical assumptions. \parm
{\leftskip 0.5in \rightskip 0.5in \noindent (iii) {In an N-world convex 
space neighborhood $I$ traced out over the time interval $[a,b],$  the NSPPM   
disturbances appear to propagate linearly. }\par}\parm 
\noindent As we proceed through this derivation$,$ other such assumptions will be 
identified.\par
The functions $q,\ \ell$ need to
 satisfy some simple mathematical 
characteristic. The best known within nonstandard analysis is the concept of 
S-continuity [8]. So$,$ where defined$,$ let $q(x)/x$ (a velocity type expression) and $\ell$
be S-continuous$,$ and $\ell$ limited (i.e. finite) at each $p \in [a,b],\ (a\!+\ {\rm at}\ a,\ b\!- \ {\rm at}\ b).$ 
From compactness$,$ $q(x)/x$ and $\ell$ are S-continuous, and 
$\ell$ is limited on $\Hyper [a,b].$  
Obviously$,$ both $q$ and $\ell$ may  have 
infinitely many totally different NSP-world characteristics of which we could 
have no knowledge. But the function $q$ represents within the NSP-world the 
distance traveled with linear units by an identifiable NSPPM disturbance. 
 It 
follows from all of this that  
for each $ t \in [a,b]$ and $t^\prime \in \monad t 
\cap \Hyper [a,b],$
$$ {{q(t^\prime)}\over{t^\prime}} - {{q(t)}\over{t}} \in \monad 0;\ \ell 
(t^\prime) - \ell (t) \in \monad 0.\eqno (3.2)$$
Expressions (3.2) give relations between nonstandard $t^\prime \in \monad 
t$ and the standard $t.$  Recall that 
if $x,\ y \in \hyperreal,$ then  $x \approx y$ iff $ x - y\in \monad 0.$ \par
 From (3.2)$,$ it follows that for each $dt \in \monad 0$ such that  $t + dt 
\in \monad t \cap \Hyper [a,b]$
$${{q(t+dt)}\over{t+dt}} \approx {{q(t)}\over {t}}, \eqno (3.3)$$ 
$$\ell( t+dt) + {{q(t+dt)}\over{t+dt}} \approx \ell (t) +{{q(t)}\over {t}}. 
\eqno (3.4)$$
\indent One important observation is necessary. The fact that the 
function $\ell$ has been evaluated at $t+dt$ is not necessary for  
(3.4) to hold for it will also hold for any $t^\prime \in \monad t$ and $\ell 
(t^\prime)$ substituted for $\ell (t+dt).$ But since we are free to choice 
any value $t^\prime \in \monad t,$ selecting particular values will allow our 
derivation to proceed to an appropriate N-world conclusion. 
From (3.4)$,$ we have that  
$$\left(\ell( t+dt) + {{q(t+dt)}\over{t+dt}}\right)dt \sim \left(\ell (t) 
+{{q(t)}\over {t}}\right)dt. 
\eqno (3.5)$$
\indent It is now that we begin our application of the concepts of classical 
Galilean composition of velocities but restrict these ideas to the NSP-world 
monadic clusters and the notion of indistinguishable effects. You will notice 
that within the NSP-world the transfer of the classical concept of equality  
of constant or average quantities is replaced by the idea of
 indistinguishable. At the moment 
$t\in [a,b]$ that 
the standard part operator is applied$,$ an effect is transmitted through the NSPPM as follows: 
\parm
{\leftskip 0.5in \rightskip 0.5in \noindent (iv) {For each $dt 
\in \monad 0$ and $t \in [a,b]$ such that $t +dt \in \Hyper [a,b],$ the 
NSP-world distance $q(t+dt) - q(t)$ (relative to $dt$) traveled by the 
NSPPM effect 
within a monadic cluster is indistinguishable for $dt$ 
from the distance produced by the Galilean composition of 
velocities.}\par}\parm
\noindent From (iv)$,$ it follows that 
$$ q(t+dt) - q(t) \sim \left(\ell (t+dt)+ {{q(t+dt)}\over{t+dt}}\right)dt.
\eqno (3.6)$$                                                                 
And from (3.5)$,$ 
$$ q(t+dt) - q(t) \sim  \left(\ell (t)+ {{q(t)}\over{t}}\right)dt.\eqno
 (3.7)$$\par
Expression (3.7) is the basic result that will lead to conclusions relative 
to the Special Theory of Relativity. In order to find out exactly what 
standard functions will satisfy (3.7)$,$ let arbitrary $t_1 \in [a,b]$ be the standard 
time at which electromagnetic propagation begins from position $F_1.$ Next$,$ let 
$q= \hyper s$ be an extended standard function 
and $s$ is continuously differentiable on $[a,b].$ Applying the definition of $\sim,$ yields 
$${{\hyper s(t + dt) -s(t)}\over{dt}} \approx \ell (t) + 
{{s(t)}\over {t}}. \eqno(3.8)$$
Note that $\ell$ is microcontinuous on $\Hyper [a,b].$  For each $t \in 
[a,b],$ the value $\ell (t)$ is limited. Hence$,$ let $\st {\ell(t)} = v(t)\in 
\real.$ From Theorem 1.1 in [3] or 7.6 in [10]$,$ $v$ is continuous on $[a,b].$ 
[See note 1 part a.] Now (3.8) may be rewritten as 
$$\Hyper {\left({{d(s(t)/t)}\over{dt}}\right)} = {{\Hyper {v(t)}}\over{t}},
\eqno (3.9)$$
where all functions in (3.9) are *-continuous on $\Hyper [a,b].$
Consequently$,$ we may apply the *-integral to both sides of (3.9). [See note 
1 part b.] Now 
(3.9) implies 
that for $t \in [a,b]$
$${{s(t)}\over{t}} =\hyper {\int_{t_1}^t}{{\Hyper v (x)}\over{x}}dx, \eqno
(3.10)$$
where$,$ for $t_1 \in [a,b],$ $s(t_1)$ has been initialized to be zero.\par
 Expression (3.10) is of 
interest in 
that it shows that although (iv) is a simplistic requirement for monadic 
clusters and the requirement that $q(x)/x$ be S-continuous is a customary 
property$,$ they do not lead to 
a simplistic NSP-world function$,$ even when view at standard NSP-world times. 
It also shows 
that the light-clock assumption was necessary in that the time represented by 
(3.10) is related to the distance traveled and unknown velocity  
of an identifiable NSPPM disturbance. It is also obvious that for pure 
NSP-world 
times the actual path of motion of such propagation  
effects is highly nonlinear in 
character$,$ although within a monadic cluster the distance $\hyper s(t + dt) - 
s(t)$ is indistinguishable from that produced by the linear-like Galilean 
composition of velocities. \par
Further$,$ it is the standard function in (3.10) that allows us to cross over 
to other monadic clusters. Thus$,$ substituting into (3.7) yields$,$ since the 
propagation behavior in all monadic clusters is identical$,$\pars  
\line{\hfil $ \hyper s(t+dt) - s(t) \sim  \left(\Hyper v (t)+  
{\left(\hyper {\int_{t_1}^t}{{\Hyper v 
(x)}/{x}}dx\right)}\right)dt,$\hfil (3.11)}\smallskip
\noindent for every $t \in [a,b],\  t+dt \in \monad t \cap \Hyper [a,b]$  \par
Consider a second standard position $F_2$ at which electromagnetic reflection 
occurs at $t_2 \in [a,b],\ t_2 > t_1,\ t_2 + dt \in \monad {t_2} \cap \Hyper [a,b].$  
Then (3.11) becomes\pars
\line{\hfil $\hyper s(t_2+dt) - s(t_2) \sim \left(\Hyper v (t_2)+  
\left(\hyper {\int_{t_1}^{t_2}}{{\Hyper v 
(x)}/{x}}dx\right)\right)dt.$\hfil (3.12)}\pars
\indent Our final assumption for monadic cluster behavior is that the classical 
ballistic property holds with respect to electromagnetic propagation.\parm 
{\leftskip 0.5in \rightskip 0.5in \noindent (v) {From the exterior 
NSP-world viewpoint$,$ at standard time $t \in [a,b],$ the velocity $\Hyper v 
(t)$ acquires an additional velocity $w$.}\par}\parm
Applying the classical statement (v), with the indistinguishable concept,  
means that the distance traveled $\hyper s(t_2 + dt) - s(t_2)$ is 
indistinguishable from $(\Hyper v (t_2) + w)dt.$ Hence$,$
$$(\Hyper v (t_2) + w)dt\sim \hyper s(t_2+dt) - s(t_2) \sim  \left(\Hyper v (t_2)+  
{\left(\hyper {\int_{t_1}^{t_2}}{{\Hyper v 
(x)}\over{x}}dx\right)}\right)dt.\eqno 
(3.13)$$                                                        
Expression (3.13) implies that 
$$\Hyper v (t_2) + w\approx \Hyper v (t_2)+  
{\left(\hyper {\int_{t_1}^{t_2}}{{\Hyper v (x)}\over{x}}dx\right)}.\eqno 
(3.14)$$                                                        
Since $\st w$ is a standard number$,$ (3.14) becomes after taking the standard 
part operator$,$                                                                 
$$ \st w=  
\St {\left(\hyper {\int_{t_1}^{t_2}}{{\Hyper v (x)}\over{x}}dx\right)}.
\eqno(3.15)$$\pars  
After reflection$,$ a NSPPM disturbance returns  to the first position $F_1$ arriving at 
$t_3  \in [a,b],\ t_1 < t_2 < t_3.$ Notice that the function $s$ does not 
appear in equation (3.15). Using the nonfavored position concept$,$ 
a reciprocal argument entails that 
$${{s_1(t_3)}\over{t_3}} = \St {\left(\hyper {\int_{t_2}^{t_3}}{{\Hyper v_1          
(x)}\over{x}}dx\right)}, \eqno(3.16)$$
$$ \st w=  
\St {\left(\hyper {\int_{t_2}^{t_3}}{{\Hyper v_1(x)}\over{x}}dx\right)},
\eqno(3.17)$$ 
where $s_1(t_2)$ is initialized to be zero. It is not assumed that $\Hyper v_1 = \Hyper v.$ \par
We now combine (3.10)$,$ (3.15)$,$ (3.16)$,$ (3.17) and obtain an 
interesting  
nonmonadic view of the relationship between distance traveled by an 
NSPPM disturbance and relative velocity. 
$$s_1(t_3) - s(t_2) = \st w (t_3 -t_2). \eqno(3.18)$$
Although reflection has been used to determine relation (3.18) and a 
linear-like interpretation involving reflection seems difficult to express$,$ there 
is a simple nonreflection analogue model for this behavior. \par
Suppose that a NSPPM disturbance is transmitted from a position $F_1,$ to 
a position $F_2.$ Let $F_1$ and $F_2$ have no NSP-world relative motion. 
Suppose that a NSPPM disturbance is transmitted from $F_1$ to $F_2$ with 
a constant velocity $v$ with the 
duration of the transmission $t^{\prime\prime} - t^\prime,$ where the path 
of motion is considered as linear. The disturbance continues linearly after it 
passes point $F_2$ but has increased during its travel through the monadic 
cluster at $F_2$ to the velocity $v+ \st w.$ The disturbance then travels 
linearly for the same duration  $t^{\prime\prime} - t^\prime.$ The linear 
difference in the two distances traveled is $w(t^{\prime\prime} - t^\prime).$
 Such results in the NSP-world should be construed only as behavior mimicked by the analogue NSPPM model. 
\par
Equations (3.10) and (3.15) show that in the NSP-world NSPPM 
disturbances propagate. Except
 for the effects of material 
objects$,$ it is assumed that in the N-world the path of motion displayed by a 
NSPPM disturbance is linear. This includes the path of motion within an 
N-world light-clock.  
We continue this derivation based upon what$,$ at present$,$ appears to be additional parameters$,$ a private NSP-world time and an NSP-world rule. 
Of course$,$  
the idea of the N-world light-clock is being used as a fixed means of identifying the different 
effects the NSPPM is having upon these two distinct worlds. A question 
yet to be answered is how can we 
compensate for differences in these two time measurements$,$ the NSP-world 
private time measurement of which we can have no knowledge and N-world 
light-clocks.\par
The weighted mean value theorem for 
integrals in nonstandard form$,$ when applied to equations  (3.15) and (3.17)$,$ states that 
there are two NSP-world times $t_a,\ t_b \in \Hyper [a,b]$ such that $t_1 \leq t_a \leq t_2 
\leq t_b \leq t_3$ and  
$$ \st w= \st {\Hyper v (t_a)} \int_{t_1}^{t_2}{{1}\over{x}}dx = 
\st {\Hyper v_1(t_b})\int_{t_2}^{t_3}{{1}\over{x}}dx.\eqno (3.19)$$
[See note 1 part c.]  
Now suppose that within the local N-world an $F_1 \to F_2,$ $F_2 \to F_1$ light-clock styled
measurement for the velocity of light using a fixed instrumentation  
yields equal quantities. (Why this is the case is established in Section 6.) Model this by (*) $\st {\Hyper v (t_a)} = \st {\Hyper v_1(t_b)}=c$ in the NSPPM. 
I point out that there are many  nonconstant *-continuous functions 
that satisfy property (*). For example$,$ certain standard nonconstant linear 
functions and nonlinear modifications of them.   
Property (*) yields  
$$ \int_{t_1}^{t_2}{{1}\over{x}}dx = 
\int_{t_2}^{t_3}{{1}\over{x}}dx.\eqno (3.20)$$ 
And solving (3.20) yields
$$\ln \left({{t_2}\over{t_1}}\right) =  \ln 
\left({{t_3}\over{t_2}}\right).\eqno (3.21)$$ 
From this one has
$$ t_2= \sqrt{t_1t_3}.  \eqno (3.22)$$ 
\par
Expression (3.22) is Prokhovnik's equation (6.3.3) in reference [7]. 
However$,$ the interpretation of this result and the others that follow 
cannot$,$ for the NSP-world$,$ be those as 
proposed by Prokhovnik. The times $t_1,\ t_2,\ t_3,$ are standard NSPPM 
times. Further$,$ it is not logically 
acceptable when considering how to measure such time in the NSP-world or 
N-world to 
consider just any mode of measurement. The mode of light velocity measurement 
must be carried out within the confines of the language used to obtain this 
derivation. Using this language$,$ a method for time calculation 
that is permissible in the N-world is the light-clock method. Any other 
described method for 
time calculation should not include significant terms from other sources. 
Time as 
expressed in this derivation is not a mystical {\it absolute} something or other. 
It is a measured quantity based entirely upon some mode of measurement. \pars

They are two major difficulties with most derivations for expressions  
used in the Special Theory. One is the above mentioned absolute time concept. 
The other is the ad hoc nonderived N-world relative velocity. 
In this case$,$ no consideration is given as to how such a relative velocity is 
to be measured so that from both $F_1$ and $F_2$ the same result would be 
obtained. It is possible to achieve such a measurement method because of the 
logical existence of the NSPPM.\pars

In a physical-like sense, the ``times'' can be considered as the numerical values recorded by single device stationary in the NSPPM. It is conceptual time in that, when events occur, then such numerical event-times ``exist.'' It is the not yet identified NSPPM properties that yield the unusual behavior indicated by (3.22). One can use light-clocks and a counter that 
indicates$,$ from some starting count$,$ the number of times the light pulse 
has traversed back and forth between the mirror and source of our light-clock. 
 Suppose that $F_1$ and  $F_2$ can coincide. When they do coincide$,$ the $F_2$ light-clock counter number that appears conceptually first after that moment can be considered to coincide with the counter number for the $F_1$ light-clock. \pars

After $F_2$ is perceived to no longer coincide with $F_1,$ a light pulse is 
transmitted from $F_1$ towards $F_2$ in an assumed linear manner.  
The ``next'' $F_1$ counter number after this event is $\tau_{11}.$  {We  
assume that the relative velocity of $F_2$ with respect to $F_1$ may have  
altered the light-clock counter numbers, compared to the count at $F_1,$ for a light-clock riding with $F_2$. The length $L$ used to define a 
light-clock is measured by the NSP-world ruler and would not be altered. Maybe the light velocity $c$, as produced by the standard part 
operator$,$ is altered by N-world relative velocity.}  Further$,$ these two  
N-world light-clocks are only located at the two positions $F_1,\ F_2,$ and  this light pulse is represented by a NSPPM disturbance. 
The light pulse is reflected back to $F_1$ by a mirror similar to the light-clock itself. The 
first counter number on the $F_2$ light-clock to appear$,$ intuitively$,$  
``after'' this reflection is approximated by $\tau_{21}.$ The $F_1$ counter number first 
perceived after the arrival of the returning light pulse is $\tau_{31}.$ \pars

 From a linear viewpoint$,$ 
at the moment of reflection$,$ denoted by $\tau_{21},$ the pulse has traveled 
an 
operational linear light-clock distance of $(\tau_{21} - \tau_{11})L.$ After 
reflection$,$ under our assumptions and nonfavored position concept$,$ 
a NSPPM disturbance would trace out the same operational linear 
light-clock distance
measured by $(\tau_{31} - \tau_{21})L.$  Thus the operational 
light-clock distance from 
$F_1$ to $F_2$ would be at the moment of operational reflection$,$ 
under our linear assumptions$,$  1/2 the sum 
of these two distances or $S_1 = (1/2) (\tau_{31} - \tau_{11})L.$ 
Now we can also determine the appropriate operational relation between these 
light-clock counter numbers for $S_1 = (\tau_{21} - \tau_{11})L.$  Hence$,$
$\tau_{31} = 2\tau_{21} -\tau_{11},$ and $\tau_{21}$ operationally behaves like an Einstein measure.\pars
After$,$ measured by light-clock counts$,$ the pulse has been received back 
to $F_1,$ a second light pulse (denoted by a second subscript of 2) is 
immediately sent to $F_2.$ Although $\tau_{31}\leq \tau_{12}$$,$ it is  
assumed that $\tau_{31}= \tau_{12}$ [See note 2.5]. 
The same analysis with new 
light-clock count numbers yields a different operational distance $S_2 = (1/2) 
(\tau_{32} -\tau_{12})L$  and $\tau_{32} = 2\tau_{22} -\tau_{12}.$  
One
 can determine the operational light-clock time intervals by considering
$\tau_{22}-\tau_{21} = (1/2)((\tau_{32} - \tau_{31}) + (\tau_{12}-\tau_{11}))$ 
and the operational linear light-clock distance difference $S_2 -S_1 = 
(1/2)((\tau_{32} - \tau_{31}) - (\tau _{12}-\tau_{11}))L.$  Since we can only 
actually measure numerical quantities as discrete or terminating numbers$,$ it 
would be empirically sound to write the N-world time intervals for these
scenarios as $t_1 = \tau_{12}-\tau_{11},t_3=(\tau_{32} - \tau_{31}).$ 
This yields the operational Einstein measure expressions in (6.3.4) of [7] as
$\tau_{22}-\tau_{21} =t_E$ and operational light-length $r_E = S_2 - S_1,$  
using our specific 
light-clock approach. This allows us to define$,$ operationally$,$ the N-world 
relative velocity as $v_E= r_E/t_E.$ [In this section$,$ the $t_1,\ t_3$ are not the same Einstein measures$,$ in form$,$ as described in [7]. But$,$ in section 4$,$ 5$,$ 6 these operational measures are used along with infinitesimal light-clock counts to obtain the exact Einstein measure forms for the time measure. This is: the $t_1$ is a specific starting count and the $t_3$ is $t_1$ plus an appropriate lapsed time.] 
\pars
Can we theoretically turn the above approximate operational approach for 
discrete N-world light-clock time  into a time continuum? Light-clocks can be 
considered from the NSP-world viewpoint. In such a case$,$ the actual NSP-world
length used to form the light-clock might be considered as a nonzero 
infinitesimal. Thus$,$ 
at least$,$ 
the numbers $\tau_{32}, \ \tau_{21}, \ \tau_{31}, \ \tau_{22}$ are 
infinite hyperreal numbers$,$ various differences would be finite and$,$ 
after taking the standard part operator$,$ all of the N-world times and lengths 
such as $t_E,\ r_E,\ S_1,\ S_2$
 should be exact and not approximate in character.
These concepts will be fully analyzed in section 6.  
Indeed$,$ as previously indicated$,$  for all of this to hold the 
velocity $c$ cannot be measured by any means. As indicated in section 6$,$ the 
actual numerical quantity $c$ as it appears in (3.22) is the standard part of 
pure NSP-world quantities. Within the N-world$,$ one obtains an ``apparent''  
constancy for the velocity of light since$,$ for this derivation$,$ 
it must be measured by means of a to-and-fro light-clock styled procedure 
with a 
fixed instrumentation.  \pars
As yet$,$ we have not discussed relations between N-world light-clock measurements 
and N-world physical laws. It should be self-evident that the assumed linearity of the light 
paths in the N-world can be modeled by the concept of projective geometry. 
Relative to the paths of motion of a light path in the NSP-world$,$ the NSPPM disturbances$,$ the
N-world path behaves as if it were a projection upon a 
plane. Prokhovnik analyzes such projective behavior and comes to 
the conclusions that in two or more dimensions the N-world light paths would 
follow the rules of hyperbolic geometry. In Prokhovnik$,$  
the equations (3.22) and the statements establishing the relations between 
the operational or exact Einstein measures $t_E,\ r_E$ and $v_E$ lead to 
the Einstein expression 
relating the light-clock determined 
 relative velocities for three 
linear positions 
having three NSP-world relative and uniform velocities $w_1,\ w_2,\ w_3.$ \pars 

In the appendix$,$ {in terms of light-clock determined Einstein measures}
and based upon the projection idea$,$ the basic Special Theory coordinate 
transformation are correctly obtained. Thus$,$ all of the NSP-world times have 
been removed from the results and even the propagation differences with 
respect to light-clock measurements. Just use light-clocks in the N-world to 
measure all these quantities in the required manner and the entire Special 
Theory is forthcoming. \par
I mention that it can be shown that  $w$ and $c$ may be measured by probes that 
are not N-world 
electromagnetic in character. Thus $w$ need not be obtained 
in the 
same manner as is $v_E$ except that N-world light-clocks would be used for 
N-world time measurements. For this reason$,$ $\st w=\omega$ is not directly related to the 
so-called textual expansion of the space within our universe. The NSPPM is not to be taken as a nonstandard translation of the Maxwell EMF equations. \par
\medskip
\leftline{\bf 4. The Time Continuum.}\par 
\medskip
With respect to models that use the classical continuum approach (i.e. 
variables are assumed to vary over such things as an interval of real 
numbers) does the mathematics perfectly measure quantities within 
nature -- quantities that cannot be perfectly measured by a human being? Or is the 
mathematics only approximate in some sense? Many would believe that if  
``nature'' is no better than the human being$,$ then classical mathematics 
is incorrect as a perfect measure of natural system behavior. However$,$ this 
is often contradicted in the limit. That is when individuals refine their 
measurements$,$ as best as it can done at the present epoch$,$ then 
the discrete human 
measurements seem to approach the classical as a limit.   
Continued exploration of this question is a 
philosophical problem that will not be discussed in this paper$,$ but it is 
interesting to model those finite things that can$,$ apparently$,$ be accomplished 
by the human being$,$ transfer these processes to the NSP-world and see what 
happens. 
For what follows$,$ when the term  ``finite''  (i.e. limited) hyperreal 
number is used$,$ since it is usually near to a nonzero real number$,$ it will 
usually refer to the ordinary  
nonstandard notion of finite except that the infinitesimals have been 
removed. This allows for the existence of finite multiplicative inverses. 
\pars  
First$,$ suppose that $t_E= \st {t_{Ea}},\ r_E= \st {r_{Ea}},\ 
S_1= \st 
{S_{1a}},\ S_2 = \st {S_{2a}}$ and each is a nonnegative real number. 
Thus $t_{Ea},\ r_{Ea},\ {S_{1a}},\  S_{2a}$ are all nonnegative  
finite hyperreal numbers. Let $L = 1/10^\omega > 0, 
\omega \in \nat^+_\infty.$ By transfer
and the result that $S_{1a},\ S_{2a},$ are  considered finite (i.e. 
near standard)$,$ then 
$S_{1a} \approx (1/2)L(\tau_{31} - \tau_{11}) \approx L
(\tau_{21} - 
\tau_{11})\Rightarrow (1/2)(\tau_{31} - \tau_{11}),\ (\tau_{21} - \tau_{11})$ 
cannot be finite. Thus$,$ by Theorem 11.1.1 [9]$,$ it can be assumed
that there exist $\eta, \gamma \in 
\nat^+_\infty$ such that
$(1/2)(\tau_{31} - \tau_{11}) =\eta,\  (\tau_{21} - \tau_{11})= \gamma.$ This 
implies that each $\tau$ corresponds to an infinite light-clock count and that 
$$\tau_{31} = 2\eta + \tau_{11},\ \tau_{21} = \gamma + 
\tau_{11}.\eqno (4.1)$$\par
In like manner$,$ it follows that
$$\tau_{32} = 2\lambda + \tau_{12},\ \tau_{22} = \delta + 
\tau_{12},\ \lambda, \delta \in \nat^+_\infty.\eqno (4.2)$$
\noindent Observe that the second of the double subscripts being 2 indicates 
the light-clock counts for the second light transmission. \pars
Now for $t_{Ea}$ to be finite requires that 
the corresponding  nonnegative $t_{1a},\ t_{3a}$ be finite. Since a different mode of conceptual time might be used in the NSP-world$,$ then there is a need for 
a number $u=L/c$ that adjusts NSP-world conceptual time to the 
light-clock count numbers. [See note 18.] By transfer of the case where  
these are real number counts$,$ this yields that $t_{3a} \approx u(\tau_{32} - 
\tau_{31}) = 2u(\lambda -\eta) +u(\tau_{12} - \tau_{11}) 
\approx 2u(\lambda -\eta) + t_{1a} $  and $t_{Ea} \approx u(\tau_{22} - 
\tau_{21}) \approx u(\delta- \gamma) + t_{1a}.$
 Hence for all of this to hold in 
the NSP-world $u(\delta -\gamma)$ must be finite or that there 
exists some $r\in \real^+$ such that $u(\delta -\gamma) \in \monad 
r.$  Let $\tau_{12} = \alpha,\ \tau_{11} = \beta.$ 
Then $t_{Ea} \approx u(\delta - \gamma) + u(\alpha - \beta)$ implies 
that $u(\alpha - \beta)$ is also finite.  \pars

The requirement that these infinite 
numbers exist in such a manner that the standard part of their products with
$L$ [resp. $u$] exists and satisfies the continuum requirements of classical mathematics 
is satisfied by Theorem 11.1.1 [9]$,$ where in that theorem $10^\omega = 1/L$ 
[resp. $1/u$]. [See note 2.]
It is obvious that the 
nonnegative numbers needed to
satisfy this theorem are nonnegative infinite 
numbers since the results are to be nonnegative and finite. 
Theorem 11.1.1 [9] allows for the appropriate $\lambda,\ \eta,\ \delta,\ \gamma$ to 
satisfy a bounding property in that we know two such numbers exist such that
$\lambda, \ \eta < 1/L^2,\ \delta,\ \gamma < 1/u^2.$ [Note: It is important to 
realize that due to this correspondence to a continuum of real numbers that 
the entire analysis as it appears in section 3 is now consistent with a mode 
of measurement. Also the time concept is replaced in this analysis with a 
``count'' concept. This count concept will be interpreted in section 8 as a 
count per some unit of time measure.] \pars
Also note that the concepts are somewhat simplified if it is assumed that
$\tau_{12} =\tau_{31}.$ In this case$,$ substitution into 4.1 yields that 
$t_{1a} \approx 2u\eta$ and $t_{3a} \approx 2u\lambda.$ Consequently$,$ $t_{Ea} = (1/2)(t_{1a} + 
t_{3a})
\approx u(\lambda+\eta).$ This predicts what is to be expected$,$ that, in this 
case$,$ the value of $t_E$ from the NSP-world viewpoint is not related to the first 
``synchronizing'' 
light pulse sent.\parm
\leftline{\bf 5. Standard Light-clocks and c.}
\medskip
I mention that the use of subparticles or the concept of the NSPPM are not 
necessary for the derivation in section 3 to hold. One can substitute for 
the NSPPM the term  ``NS-substratum'' or the like and for the  term
``monadic cluster'' of possible subparticles  just the concept of a 
``monadic neighborhood.''  It is not necessary that one assume that the 
NS-substratum is composed of subparticles or any identifiable entity$,$ only  
that NSPPM  transmission of such radiation behaves in the 
simplistic manner stated.\pars
It is illustrative to show by a diagram of simple light-clock counts how this 
analysis actually demonstrates the two different modes of  propagation$,$ the NSP-world
 mode and the different mode when viewed from 
the N-world. In general$,$ $L$ is always fixed and for the following analysis
and$,$ for this particular scenario$,$ inf. light-clock $c$ 
may change. This process of using N-world light-clocks to approximate
the relative velocity should only be done once due to the necessity of 
``indexing'' the light-clocks when $F_1$ and $F_2$ coincide. 
In the following diagram$,$ the numbers represent actual 
light-clock count numbers as perceived in the N-world.
 The first column are those recorded at $F_1,$ the 
second column those required at $F_2.$  The arrows 
and the numbers above them represent our $F_1$ comprehension of 
what happens when the transmission is considered to take place in the 
N-world.   
The Einstein measures are only for the $F_1$ position.  
 \parm
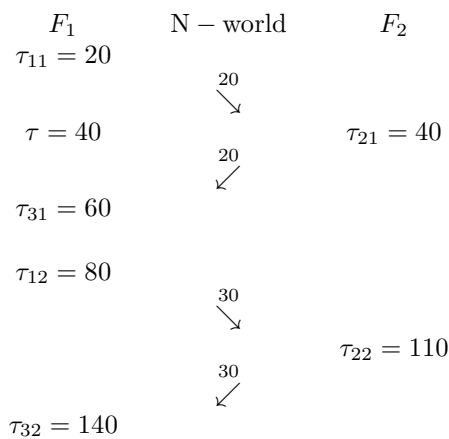
$$ \matrix{F_1&&{\rm N-world}&&F_2\cr
\tau_{11} = 20&&&&&\cr
&&{\buildrel 20 \over \searrow}&&\cr
\tau = 40 &&&&\tau_{21} = 40\cr
&&{\buildrel 20 \over \swarrow}&&\cr
\tau_{31} = 60&&&&&\cr
&&&&&&\cr
\tau_{12} = 80&&&&&\cr
&&{\buildrel 30 \over \searrow}&&\cr
&&&&\tau_{22} = 110\cr
&&{\buildrel 30 \over \swarrow}&&\cr
\tau_{32} =140&&&&&\cr}$$   
\medskip 
Certainly$,$ the above diagram satisfies the required light-clock count 
equations.  
The only light-clock counts that actually are perceivable are those at $F_1.$
And$,$ for the  transformation equations$,$ the scenario is altered. 
When the Special Theory transformation equations are obtained$,$ two distinct N-world observers are used and a third 
N-world distinct fundamental position. All light-clock counts made at each of 
these three positions are entered  into the  appropriate expressions 
for the Einstein measures {\bf as obtained for each individual position.}\par 
\bigskip
\leftline{\bf 6. Infinitesimal Light-clock Analysis.}
\medskip
In the originally presented Einstein derivation$,$ time and length are 
taken as absolute time and length. It was previously 
pointed out that this 
assumpt yields logical error. The scientific
  community extrapolated 
the language used in the derivation$,$ a language stated only in terms of 
light propagation behavior$,$ without logical reason, to the ``concept'' of 
Newtonian absolute time and length. Can the actual meaning of the 
``time'' and ``length'' expressed in the Lorentz transformation be 
determined?\parm

In what follows$,$ a measure by light-clock counts is used to analyze 
the classical transformation as derived in the Appendix-A and, essentially, such ``counts'' will replace conceptional time. [See note 1.5] The superscripts 
indicate the counts associated with the light-clocks$,$ the Einstein measures$,$ 
and the like$,$ at the positions $F_1, \ F_2.$ The $1$ being the light-clock measures at $F_1$ 
for a light pulse event from $P,$ the $2$ for the light-clock measures at the $F_2$ 
for the same light pulse event from $P,$ and the 3 for the light-clock measures and its corresponding Einstein measures at $F_1$ for the  velocity of $F_2$ relative to $F_1.$ The 
NSP-world measured angle$,$ assuming linear projection due to the constancy of 
the velocities$,$ from $F_1$ to the light pulse event from $P$ is $\theta,$ and that from $F_2$ to $P$ is an exterior angle $\phi.$  \pars

The expressions for our proposes are $x^{(1)}_E = v^{(1)}_Et^{(1)}_E\cos 
\theta,\  x^{(2)}_E = -v^{(2)}_Et^{(2)}_E\cos 
\phi.$ [Note: The negative is required since $\pi/2 \leq \phi \leq \pi$ and use of the customary coordinate systems.] In all that follows$,$ $i$ 
varies from 1 to 3. We investigate what happens when the standard model is 
now embedded back again into the {\it non-infinitesimal finite} NSP-world. All of the  
``coordinate'' transformation equations are in the Appendix and they  actually only involve 
$\omega_i/c.$ These equations are interpreted in the NSP-world. 
But  
as far as the light-clock counts are concerned$,$  their appropriate 
differences   
are only infinitely near to a standard number. The appropriate expressions are 
altered to take this into account. For simplicity in notation$,$ it is again 
assumed that   ``immediate'' in the light-clock count process means $\tau^{(i)}_{12} 
= \tau^{(i)}_{31}.$ [See note 3.] Consequently$,$  
$t^{(i)}_{1a} 
\approx 2u\eta^{(i)},\ t^{(i)}_{3a} 
\approx 2u\lambda^{(i)},\ \eta^{(i)},\lambda^{(i)}\in \nat^+_\infty.$ Then
$$t^{(i)}_{Ea}\approx  u(\lambda^{(i)}+\eta^{(i)}),\ \lambda^{(i)},\eta^{(i)}\in \nat^+_\infty. \eqno (6.1)$$  \par
Now from our definition $r^{(i)}_E \approx L(\lambda^{(i)} - 
\eta^{(i)}),\ (\lambda^{(i)} - 
\eta^{(i)})\in \nat^+_\infty.$  Hence, since all of the numbers to which {\tt st} is applied 
are nonnegative and finite and $\st {v^{(i)}_{Ea}}\ \st {t^{(i)}_{Ea}} = \st 
{r^{(i)}_{Ea}},$ it follows that
$$v^{(i)}_{Ea}\approx L{{(\lambda^{(i)} - 
\eta^{(i)})}\over{u(\lambda^{(i)}}+\eta^{(i)})}.\eqno (6.2)$$ 
Now consider a set of two
4-tuples $$(\st {x^{(1)}_{Ea}},\st {y^{(1)}_{Ea}},\st {z^{(1)}_{Ea}},
\st {t^{(1}_{Ea})},$$
$$(\st {x^{(2)}_{Ea}},\st {y^{(2)}_{Ea}},\st {z^{(2)}_{Ea}},\st {t^{(2)}_{Ea}}),$$ where 
they are viewed as Cartesian coordinates in the NSP-world. First$,$ we have
$\st {x^{(1)}_{Ea}} = \st {v^{(1)}_{Ea}}\st {(t^{(1)}_{Ea}}\st {\hyper \cos \theta},\ 
\st {x^{(2)}_{Ea}} = \st {v^{(2)}_{Ea}}\st {t^{(2)}_{Ea}}\st {\hyper \cos \phi}.$
Now suppose the local constancy of  $c$. The N-world Lorentz transformation 
expressions are
$$\st {t^{(1)}_{Ea}} =  {\beta_3}(\st {t^{(2)}_{Ea}} + \st {v^{(3)}_{Ea}}\st 
{x^{(2)}_{Ea}}/c^2),$$ 
$$\st {x^{(1)}_{Ea}}= {\beta_3}(\st {x^{(2)}_{Ea}} + \st {v^{(3)}_{Ea}}\st 
{t^{(2)}_{Ea}}),$$
where $\beta_3 = \st {(1 - (v^{(3)}_{Ea})^2/c^2)^{-1/2}}.$ 
Since $L(\lambda^{(i)} - \eta^{(i)}) \approx cu(\lambda^{(i)} - \eta^{(i)})$$,$ 
the finite character of $L(\lambda^{(i)} - \eta^{(i)}),\ u(\lambda^{(i)} - 
\eta^{(i)})$ yields that $c= \st {L/u}$ [See note 8].  When transferred to the NSP-world 
with light-clock  counts$,$ substitution yields
$$t^{(1)}_{Ea}\approx u(\lambda^{(1)}+\eta^{(1)})
\approx \beta[u(\lambda^{(2)}+\eta^{(2)}) -u(\lambda^{(2)}+\eta^{(2)})K^{(3)}K^{(2)}\hyper \cos 
\phi],\eqno (6.3)$$
where $K^{(i)} = (\lambda^{(i)} - \eta^{(i)})/(\lambda^{(i)}+\eta^{(i)}),\  \beta 
= (1 - (K^{(3)})^2)^{-1/2}.$ \par
For the  ``distance''
               transformation$,$ we have
$$x^{(1)}_{Ea} \approx L(\lambda^{(1)} - \eta^{(1)})\hyper 
\cos \theta 
\approx$$ 
$$\beta(-L(\lambda^{(2)} - \eta^{(2)})
\hyper \cos \phi +{{L(\lambda^{(3)} - 
\eta^{(3)})}\over{u(\lambda^{(3)}+\eta^{(3)})}}u(\lambda^{(2)}+\eta^{(2)})).\eqno (6.4)$$
Assume 
in the NSP-world 
that $\theta \approx \pi/2,\ \phi \approx \pi.$ Consequently$,$ 
substituting into 6.4 yields
$$-L(\lambda^{(2)} - \eta^{(2)})\approx
{{L(\lambda^{(3)} - 
\eta^{(3)})}\over{u(\lambda^{(3)}+\eta^{(3)})}}u(\lambda^{(2)}+\eta^{(2)}).\eqno (6.5)$$\par 
Applying the finite property for these numbers$,$ and, for this scenario, taking into account the different modes of the corresponding light-clock measures, yields
$$ {{L(\lambda^{(3)} - 
\eta^{(3)})}\over{u(\lambda^{(3)}+\eta^{(3)})}}\approx {{-L(\eta^{(2)} - 
\lambda^{(2)})}\over{u(\lambda^{(2)}+\eta^{(2)})}} \Rightarrow v^{(3)}_{Ea} \approx - v^{(2)}_{Ea}. 
\eqno (6.6)$$
Hence$,$ $\st {v^{(3)}_{Ea}} =-\st {v^{(2)}_{Ea}}.$ [Due to the coordinate-system  selected, these are directed velocities.]  This predicts that$,$ in the 
N-world$,$ the light-clock determined relative velocity of $F_2$
as measured from the $F_1$ and $F_1$ as measured from the $F_2$ positions would be 
the same if these 
special infinitesimal light-clocks are used. If noninfinitesimal 
N-world light-clocks are used$,$ then the values will be 
approximately the same and  equal in the limit. \par 
Expression 6.4 relates the light-clock counts relative to the measure of the 
to-and-fro paths of light transmission. By not  substituting for 
$x^{(2)}_{Ea},$ 
it is easily seen that $x^{(2)}_{Ea}\approx LG,$ where $G$ is an expression 
written entirely in terms of various light-clock count numbers. This implies 
that 
the so-called 4-tuples $(\st {x^{(1)}_{Ea}}, \st {y^{(1)}_{Ea}},$  
$\st {z^{(1)}_{Ea}},$ $\st {t^{(1}_{Ea})},$
$(\st {x^{(2)}_{Ea}},$$\st {y^{(2)}_{Ea}},$ $\st {z^{(2)}_{Ea}},
\st {t^{(2)}_{Ea}})$ 
are not the absolute Cartesian 
type coordinates determined by  Euclidean geometry and used to model 
Galilean dynamics. These coordinates are dynamically 
determined by the behavior of electromagnetic radiation within the 
N-world. Indeed$,$ in [7]$,$ the analysis within the  (outside of the 
monadic clusters) that leads to Prokhovnik's conclusions is only relative to 
electromagnetic propagation and is done by pure number Galilean dynamics. 
Recall that the 
monadic cluster analysis is also done by Galilean dynamics.  
\par    
In general$,$ when it is claimed that  ``length contracts'' with respect to 
relative velocity the  ``proof'' is stated as follows: $x^\prime =
\st {\beta}(x + vt); \ {\overline{x}}^\prime = \st {\beta}({\overline{x}} + 
\overline{v}\overline{t}).$ Then these two 
expressions are subtracted.  Supposedly$,$ this yields $                            
{\overline{x}}^\prime -x^\prime= \st {\beta}(\overline{x} - x)$ since its assumed 
that $\overline{v}\overline{t} = vt.$
 For defined coordinates $\overline{x}_E^{(i)},\ x_E^{(i)},\ i = 1,2,$ a more complete expression would be 
$${\overline{x}}^{(1)}_E -x^{(1)}_E= \st {\beta}((\overline{x}^{(2)}_E - x^{(2)}_E)+ 
(\overline{v}^{(3)}_E\overline{t}^{(2)}_E - v^{(3)}_Et^{(2)}_E)).\eqno (6.7)$$\par
In this particular analysis$,$ it has been assumed that all NSP-world relative 
velocities $\omega_{i}, \overline{\omega_i} \geq 0.$ To obtain the classical 
length contraction expression$,$ let $\omega_{i}= \overline{\omega_i},\ 
i=1,2,3.$  Now this  implies that 
$\overline{\theta} = \theta,\ \overline{\phi} = \phi$ as they appear in the 
velocity figure on page 52 and that  
$${\overline{x}}^{(1)}_E -x^{(1)}_E= \st {\beta}(\overline{x}^{(2)}_E - 
x^{(2)}_E).\eqno (6.8)$$\par
The difficulty with this expression has been its interpretation.
Many modern treatments of Special Relativity [6] argue that (6.8) has no 
physical meaning. But in these arguments it is assumed that 
${\overline{x}}^{(1)}_E -x^{(1)}_E$ means ``length'' in the Cartesian 
coordinate sense as related to Galilean dynamics. As pointed out$,$ such a   
physical meaning is not  
the case. Expression (6.8) is a relationship between light-clock counts and$,$ 
in general$,$ displays properties of electromagnetic propagation within the 
N-world. Is there a difference between the right and left-hand sides of 6.8 
when viewed entirely from the NSP-world. First$,$ express 6.8 as
$\overline{x}^{(1)}_E -x^{(1)}_E= \st \beta\overline{x}^{(2)}_E - 
\st \beta x^{(2)}_E.$ In terms of operational light-clock counts$,$ this expression becomes
$$L(\overline{\lambda}^{(1)}\hyper {\cos {\theta}} - 
\overline{\eta}^{(1)}
\hyper {\cos {\theta}}) - L(\lambda^{(1)}\hyper {\cos \theta} - 
\eta^{(1)}\hyper {\cos \theta})
\approx\eqno (6.9)$$ 
$$L(\overline{\lambda}^{(2)}\beta{\vert \hyper {\cos {{\phi}}}\vert} - 
\overline{\eta}^{(2)}\beta
{\vert \hyper {\cos {\phi}\vert })} -  L(\lambda^{(2)}\beta {\vert\hyper {\cos 
\phi\vert}} 
- \eta^{(2)}\beta
{\vert \hyper {\cos \phi\vert}}),$$ 
where finite $\beta = (1 - (K^{(3)})^2)^{-1/2}$ and $\vert \cdot \vert$ is 
used so that the Einstein velocities are not directed numbers and the 
Einstein distances  are comparable. Also as  long as $\theta,\ \phi$ satisfy 
the velocity figure on page 45$,$ then (6.9) is independent of the specific angles
chosen in the N-world since in the N-world expression (6.8) no angles 
appear relating the relative 
velocities. That is, the velocities are not vector quantities in the N-world$,$ 
but scalars.\par

Assuming the nontrivial case that $\theta \not\approx 
\pi/2, \ \phi \not\approx \pi,$ we have from Theorem 11.1.1 [9]
that there exist $\overline{\Lambda}^{(i)},\ \overline{N}^{(i)},\ 
\Lambda^{(i)},\ N^{(i)} \in \nat_\infty,\ i=1,2$ such that
$\hyper  {\cos {\theta}}\approx \overline{\Lambda}^{(1)}/
\overline{\lambda}^{(1)} \approx 
\overline{N}^{(1)}/\overline{\eta}^{(1)}\approx$
${\Lambda}^{(1)}/
{\lambda}^{(1)} \approx {N}^{(1)}/{\eta}^{(1)},$
$\beta\vert \hyper {{\cos {\phi}}}\vert\approx \overline{\Lambda}^{(2)}/
\overline{\lambda}^{(2)} \approx \overline{N}^{(2)}/\overline{\eta}^{(2)}$
$\approx {\Lambda}^{(2)}/
{\lambda}^{(2)} \approx {N}^{(1)}/{\eta}^{(2)}.$  Consequently$,$ using the 
finite character of these quotients and the finite character of 
$L(\overline{\lambda}^{(i)}),\ L(\overline{\eta}^{(i)}),\ 
L({\lambda}^{(i)}),\ L({\eta}^{(i)}), \  i = 1,2,$ the general 
three body NSP-world
view 6.9 is
$$L(\overline{\Lambda}^{(1)} -\overline{N}^{(1)})-L(\Lambda^{(1)}-N^{(1)}) 
= L\Gamma^{(1)}\approx$$ $$ L\Gamma^{(2)}_1 =L(\overline{\Lambda}^{(2)} -\overline{N}^{(2)})-L(\Lambda^{(2)}-
N^{(2)}). \eqno (6.10)$$\par
The obvious interpretation of 6.10 from the simple NSP-world 
light propagation viewpoint is displayed by taking the standard part of 
expression 6.10.\pars

$$\st {L(\overline{\Lambda}^{(1)} -\overline{N}^{(1)})}-\st {L(\Lambda^{(1)}-
N^{(1)})}=\st {L\Gamma^{(1)}} 
=$$ $$\st {L\Gamma^{(2)}_1}= \st {L(\overline{\Lambda}^{(1)} -\overline{N}^{(1)})}-\st {L(\Lambda^{(1)}-
N^{(1)})}. \eqno (6.11)$$\par
This is the general view as to the equality of the standard NSP-world 
distance traveled by a light pulse moving to-and-fro within a light-clock as 
used to measure at $F_1$ and 
$F_2,$ as viewed from the NSPPM only$,$ the occurrence of the light pulse event from $P$. In order to interpret 6.9 for the N-world and a single NSP-world 
relative velocity$,$ you consider additionally that $\omega_1 
=\omega_2=\omega_3.$  Hence$,$
$\theta = \pi/3$ and 
correspondingly ${\phi} = 2\pi/3.$ In this case$,$  
 $\beta$ is unaltered and since $\cos \pi/3,\ \cos 2\pi/3$ are nonzero 
and finite$,$  
 6.9 now yields
$$\st {L(\overline{\lambda}^{(1)} -\overline{\eta}^{(1)})}-\st {L(\lambda^{(1)}-
\eta^{(1)})} 
=$$ $$ \st {\beta}(\st {L(\overline{\lambda}_1^{(2)} -\overline{\eta}_1^{(2)})}-
\st {L(\lambda_1^{(2)}-
\eta_1^{(2)})})\Rightarrow$$
$$(\st{L\overline{\lambda}^{(1)}} -\st {L\overline{\eta}^{(1)}})-(\st 
{L\lambda^{(1)}}-
\st {L\eta^{(1)}}) 
=$$ $$ \st {\beta}((\st {L\overline{\lambda}_1^{(2)}} -\st 
{L\overline{\eta}_1^{(2)}})-
(\st {L\lambda_1^{(2)}}-
\st {L\eta_1^{(2)}})). \eqno (6.12)$$
Or\vfil\eject
$$\st { L(\overline{\lambda}^{(1)} -\overline{\eta}^{(1)})- L(\lambda^{(1)}-
\eta^{(1)})} =$$                                               
$$\st { L[(\overline{\lambda}^{(1)} -\overline{\eta}^{(1)})- (\lambda^{(1)}-
\eta^{(1)})] }=$$
$$\st { L{\Pi}^{(1)} }=\st {\beta}\st {L{\Pi}_1^{(2)} }=\st 
{\beta L{\Pi}_1^{(2)} }=\eqno (6.13)$$ 
$$\st { L[(\overline{\lambda}^{(1)} -\overline{\eta}^{(1)})- (\lambda^{(1)}-
\eta^{(1)})] }=$$
$$
\st {\beta L[(\overline{\lambda}_1^{(2)} -\overline{\eta}_1^{(2)})-
(\lambda_1^{(2)}-
\eta_1^{(2)})]}. $$
In order to obtain the so-called  ``time dilation'' expressions$,$ follow the 
same procedure as above. Notice$,$ however$,$ that (6.3) leads to a contradiction unless 
$$u((\overline{\lambda}^{(1)} + \overline{\eta}^{(1)})- (\lambda^{(1)}+\eta^{(1)})) \approx\beta 
u((\overline{\lambda}^{(2)}+\overline{\eta}^{(2)}) - (\lambda^{(2)}+\eta^{(2)})).\eqno (6.14)$$ 
It is interesting$,$ but not surprising$,$ that this procedure yields (6.14) without hypothesizing a relation between the $\omega_i,\ i= 1,2,3$ and implies that the timing infinitesimal light-clocks are the fundamental constitutes for the analysis.
In the NSP-world$,$ 6.14 can be re-expressed as
$$u((\overline{\lambda}^{(1)}+\overline{\eta}^{(1)}) - (\lambda^{(1)}+\eta^{(1)})) \approx 
u(\overline{\lambda}_2^{(2)} - \lambda_2^{(2)}).\eqno (6.15)$$  
\noindent Or
$$\st {u((\overline{\lambda}^{(1)}+\overline{\eta}^{(1)})} =\st 
{u\Pi_2^{(1)}}=$$ $$ 
\st{u\Pi_3^{(2)}}= \st {u(\overline{\lambda}_2^{(2)} - 
\lambda_2^{(2)})}.\eqno (6.16)$$ 
[See note 4.] 
From the N-world$,$ the expression becomes$,$ taking the standard part operator$,$
$$\st {u(\overline{\lambda}^{(1)}+\overline{\eta}^{(1)})} - \st {u(\lambda^{(1)}+\eta^{(1)})}=
$$ $$
\st{\beta}(\st{u(\overline{\lambda}^{(2)}+\overline{\eta}^{(2)})} - \st {u(\lambda^{(2)}+\eta^{(2)}))}.
\eqno (6.17)$$\par
\noindent Or
$$\st {u\Pi_2^{(1)}}=\st{\beta}\st {u\Pi_4^{(2)}}= \st 
{ \beta u\Pi_4^{(2)}}=$$
$$\st {u((\overline{\lambda}^{(1)}+\overline{\eta}^{(1)}) - (\lambda^{(1)}+\eta^{(1)}))} =    
\st {\beta u[(\overline{\lambda}^{(2)}+\overline{\eta}^{(2)}) - (\lambda^{(2)}+\eta^{(2)})]}.\eqno (6.18)$$
Note that using the standard part operator in the above expressions$,$ yields  
continuum time and space coordinates to which the calculus can now be applied. 
However$,$ the time and space measurements are not to be made with respect to an 
 universal (absolute) clock or ruler. The measurements are relative to electromagnetic 
propagation. The Einstein time and length are not the NSPPM time and length$,$ 
but rather they are concepts that incorporate a mode of measurement into 
electromagnetic field theory. This mode of measurement follows from the one wave property used for Special Theory scenarios, the property that, in the N-world, the propagation of a photon do not take on the velocity of its source. It is this that helps 
clarify properties of the NSPPM. 
 Expressions such as (6.13)$,$ (6.18) will be interpreted in the  next 
sections of this paper.\par
\medskip 
 \leftline{\bf 7. An Interpretation.}
\medskip
In each of the expressions  $(6.i),\ i=10,\ldots, 18$ 
the infinitesimal numbers $L,\ u$ are unaltered. If this is the case, then the light-clock counts would appear to be altered. As shown in Note [2], alteration of $c$ can be represented as alterations that yield infinite counts. Thus, in one case, you have a specific infinitesimal $L$ and for the other infinitesimal light-clocks a different light-clock $c$ is used. But, $L/u = c.$  Consequently the only 
alteration that takes place in N-world expressions $(6.i),
\ i=12,13,17,18$ is the infiniteimal light-clocks that need to be employed. This is exactly what (6.13) 
and (6.18) state if you consider it written as say, $(\beta L)\  \cdot$ rather than $L (\beta \ \cdot)$. Although these are external 
expressions and cannot be  ``formally'' transferred back to the N-world$,$ the 
methods of infinitesimal modeling require the concepts of  ``constant'' and 
``not constant'' to be preserved. \par 

These N-world expressions can be 
re-described in terms of N-world approximations. Simply substitute $\doteq$ for 
$=,$ a nonzero real $d$ [resp. $\mu$] for $L$ [resp. $u$] and 
real natural numbers for each 
light-clock count in equations $(6.i)$,$
\ i=12,17.$ Then for a particular $d$ [resp. $\mu$] any change 
in the light-clock
measured relative velocity $v_E$ would dictate a change in the 
the light-clocks used. 
Hence$,$ the N-world need not be concerned with the idea that  ``length''
contracts but rather it is the required light-clocks change. It is the required change in infiniteimal light-clocks that lead to real physical changes in behavior as such behavior is compared to a standard behavior. {\it But$,$ in many cases$,$  the use of light-clocks is 
not intended to 
be a literal use of such instruments.} For certain scenarios$,$ light-clocks are 
to be considered as  {\it analog 
models} that incorporate  electromagnetic energy properties. [See note 18, first paragraph.]  
\par 
The analysis given in the section 3 is done to discover 
a general property for the transmission of electromagnetic radiation. 
It is clear that property (*) does not require that the  measured velocity of 
light be a universal constant. All that is needed is that for the two 
NSP-world  times $t_a,\ t_b$ that $\st {\ell (t_a)} = \st {\ell_1 (t_b)}.$
 This means that all that is required for the most basic aspects of the 
 Special Theory 
to hold is that at two NSP-world times in the $F_1 \to F_2,\ F_2 \to F_1$  reflection 
process $\st {\ell (t_a)} = \st {\ell_1 (t_b)}$$,$ $t_a$ a time 
during the transmission prior to reflection and $t_b$ after 
reflection.  If $\ell,\ \ell_1$ are nonstandard extensions of standard functions 
$v,\ v_1$ continuous on $[a,b]$$,$ then given any $\eps \in \real^+$ 
there is a $\delta$ 
such that
for each $t,\ t^\prime \in [a,b]$ such that $\vert t - t^\prime \vert < 
\delta$ it follows  that 
$\vert v(t)-v(t^\prime) \vert < \eps/3 $ and 
$\vert v_1(t)-v_1(t^\prime) \vert < \eps/3.$ Letting $t_3 -t_1 < 
\delta,$ then $\vert t_a - t_b \vert < \delta.$ Since $\Hyper v(t_a) = 
\ell(t_a) \approx \Hyper v_1(t_b) = \ell_1(t_b),$ *-transfer implies
$\vert \Hyper v(t_2) - \Hyper v_1(t_2) \vert < \eps.$ [ See note 5.] 
Since $t_2$ is a 
standard number$,$ $\vert  v(t_2) - v_1(t_2) \vert < \eps$  implies that 
$v(t_2) = v_1(t_2).$ Hence$,$ in this case$,$ the two functions $\ell,\ \ell_1$ do 
not differentiate between the velocity $c$ at $t_2.$ But $t_2$ can be considered 
an arbitrary (i.e. NSPPM) time such that $t_1 < t_2 < t_3.$ {\bf This does not require  
$c$ to be the same 
for all cosmic times} only that $v(t) = v_1(t),\ t_1 < t < t_3.$ \par 
{\it The restriction that $\ell,\ \ell_1$ are extended standard functions 
appears necessary for our derivation.} Also$,$ this analysis is not 
related to what $\ell$ may be for a stationary laboratory. In the case of 
stationary $F_1,\ F_2,$ then the integrals are zero in equation (19) of 
section 3. The 
easiest thing to do is to simply postulate that $\st {\Hyper v (t_a)} $ is a 
universal constant. This does not make such an assumption correct. 
\par
One of the properties that will allow the Einstein velocity transformation 
expression to be derived is the {\it equilinear} property. {\bf This property 
is 
weaker than the $c$ = constant property for light propagation.} 
Suppose that you have within the NSP-world three observers 
$F_1,\ F_2,\ F_3$ that are linearly related. Further$,$ suppose that
$w_1$ is the NSP-world velocity of $F_2$ relative to $F_1$ and 
$w_2$ is the NSP-world velocity of $F_3$ relative to $F_2.$ It is assumed 
that for this nonmonadic cluster situation$,$ that Galilean dynamics also 
apply and that $\st {w_1} + \st {w_2} = \st {w_3}.$ Using the description  
for light propagation as given in section 3$,$ let $t_1$ be the 
cosmic time when a light pulse leaves $F_1$$,$ $t_2$ when it  ``passes''  
$F_2,$  and $t_3$ the cosmic time when it arrives at $F_3.$\par
From equation (3.15)$,$ it follows that
$$\st {w_1} = \st {\Hyper v_1(t_{1a})} 
\St {\left(\hyper {\int_{t_1}^{t_2}}{{1}\over{x}}dx\right)}+ 
[\st {w_2} =] \st {\Hyper v_2(t_{2a})} 
\St {\left(\hyper {\int_{t_2}^{t_3}}{{1}\over{x}}dx\right)}=$$ 
$$\st {w_3} = \st {\Hyper v_3(t_{3a})} 
\St {\left(\hyper {\int_{t_1}^{t_3}}{{1}\over{x}}dx\right)}.\eqno 
(7.1)$$ 
If $\st {\Hyper v_1(t_{1a})}=\st {\Hyper v_2(t_{2a})}=\st {\Hyper v_3(t_{3a})},$ then we 
say that the velocity functions $\Hyper v_1, \Hyper v_2, \Hyper v_3$ are {\it equilinear.} 
The constancy of $c$ implies equilinear$,$ but not conversely.  
In either case$,$ functions such as $\Hyper v_1$ and $\Hyper v_2$ need not be the same 
within a stationary laboratory after interaction. \par
Experimentation indicates that electromagnetic propagation does  
``appear'' to behave in the N-world in such a way that it does not accquire the velocity of the source. 
The light-clock analysis is consistent with the following speculation.
{\bf Depending upon the scenario$,$
the uniform velocity 
yields an effect via interactions with the subparticle field (the \underbar{NSPPM}) that uses a photon particle behavioral model.  
This is termed the \underbar{(emis)} effect.}
Recall that a ``light-clock''  can 
be considered as an analog model for the most basic of the 
electromagnetic properties. On the other hand$,$ only those experimental  methods 
that replicate or are equivalent to the methods of Einstein measure
would be relative to the Special Theory. This is one of the basic logical 
errors in theory application. The experimental language must be related to 
the language of the derivation. The concept of the light-clock$,$ linear paths 
and the like are all intended to imply NSPPM interactions. Any 
explanation for experimentally verified Special Theory effects should be stated in 
such a language and none other. 
I also point out that there are no paradoxes in this derivation for you cannot 
simply  ``change your mind'' with respect to the NSPPM. For example$,$ 
an observer is either in motion or 
not in motion$,$ and  not both {\bf with respect to the NSPPM.} \pars   
\leftline{\bf 8. A Speculation and Ambiguous Interpretations}\pars
Suppose that the correct principles of infinitesimal modeling were 
known prior to the M-M (i.e. Michelson-Morley) experiment. Scientists would 
know that the (mathematical) NSPPM is not an N-world entity. They would 
know that they could have very little knowledge as to the refined workings of 
this NSP-world NSPPM since $\approx$ is not an $=.$ They would have been 
forced to accept the statement of Max Planck that ``Nature does not allow 
herself to be exhaustively expressed in human thought.''[{\it The Mechanics of 
Deformable Bodies$,$ Vol. II$,$ Introduction to Theoretical Physics}$,$ Macmillian$,$ 
N.Y. (1932)$,$p. 2.]\par
Further suppose$,$ that human comprehension was advanced enough so that all 
scientific experimentation always included a theory of measurement.  
The M-M experiment would then have been performed to 
learn$,$ if possible$,$ more about this NSP-world NSPPM. When a null finding 
was obtained then a derivation such as that in section 3 might have been 
forthcoming. Then the following two expressions would have emerged from the 
derivation.\par         
The Einstein method for measurement - the ``radar'' method - is used (see A3, p. 52) to determining the relative velocity of the moving light-clock. Using Appendix-A equations (A14), let $P$ correspond to $F_2$. Then $\theta = 0, \ \phi = \pi/2.$ Since, $x^{(2)} = 0$ from page 54, then $F_2$ is the $s$-point Hence, $t^2_E =t^{(2)}$. The superscript and subscript $s$ represents local measurements about the $s$-point, using various devices, for laboratory standards (i.e. standard behavior) and using infinitesimal light-clocks or approximating devices such as atomic-clocks. [Due to their construction atomic clocks are effected by relativistic motion and gravitational fields approximately as the infinitesimal light-clock's counts are effected.] Superscript or subscript $m$ indicates local measurements, using the same devices, for an entity considered at the $m$-point in motion relative to the $s$-point, where Einstein time and distance via the radar method as registered at $s$ are used to investigate $m$-point behavior. For example, $m$-point time is measured at the $s$-point via infinitesimal  light clock and the radar method and this represents time at the $m$-point. To determine how physical behavior is being altered, the $m$ and $s$-measurements are compared. Many claim that you can replace each $s$ with $m$, and $m$ with $s$ in what follows. This may lead to various controversies which are elimianted in part 3.  A specific interpretation of
$$\st {\beta}^{-1}({\overline{t}}^{(s)} -t^{(s)})= 
\overline{t}^{(m)}_E - 
t^{(m)}_E \eqno (8.1)$$
or the corresponding  
 $$\st {\beta}^{-1}({\overline{x}}^{(s)} -x^{(s)})= 
\overline{x}^{(m)}_E - 
x^{(m)}_E \eqno (8.2)$$  
seems necessary. However, (8.2) is unnecessary since $v_E(\st {\beta}^{-1}({\overline{t}}^{(s)} -t^{(s)}))= v_E(
\overline{t}^{(m)}_E - 
t^{(m)}_E)$ yields (8.2), which can be used when convienient. Thus, only the infinitesimal light-clock ``time'' alterations are significant. Actual length as measured via the radar method is not altered. It is the clock counts that are altered. \par

\underbar{If}$,$ in (8.2)$,$ which is employed for convenience, ${\overline{x}}^{(s)} - x^{(s)}= U^s$ (note that $x^{(s)} = v_Et^{(s)}$ etc.) is 
interpreted as ``any'' standard unit  
for length measurement at the $s$-point and 
$\overline{x}^{(m)}_E - 
x^{(m)}_E)= U^m$ the same ``standard'' unit for length measurement in a system 
moving with respect to the NSPPM (without regard to direction)$,$ 
 then for equality to take place the unit of measure $U^m$ may seem to be altered 
in the moving system. Of course$,$ it would have been immediately realized 
that the error in this last statement is that 
$U^s$ is ``any'' unit of measure. Once again$,$ the error in these two statements is the term
``any.'' (This problem is removed by application of $(14)_a$ or $(14)_b$ p. 60.)\par

\underbar{If}$,$ in (8.2)$,$ which is employed for convenience, ${\overline{x}}^{(s)} - x^{(s)}= U^s$ (Note that $x^{(s)} = v_Et^{(s)}.$) is 
interpreted as ``any'' standard unit  
for length measurement at the $s$-point and 
$\overline{x}^{(m)}_E - 
x^{(m)}_E)= U^m$ the same ``standard'' unit for length measurement in a system 
moving with respect to the NSPPM (without regard to direction)$,$ 
 then for equality to take place the unit of measure $U^m$ may seem to be altered 
in the moving system. Of course$,$ it would have been immediately realized 
that the error in this last statement is that 
$U^s$ is ``any'' unit of measure. Once again$,$ the error in these two statements is the term
``any.'' (This problem is removed by application of $(14)_a$ or $(14)_b$ p. 60.)\par

 Consider 
experiements such as the M-M$,$ Kennedy-Thorndike and many others.  
When viewed from the wave state$,$ the  
interferometer measurement  technique is  determined completely by a 
light-clock type process -- the \underbar{number} of light waves in the 
linear path.  We need to use $L_{sc}^m,$ a scenario associated 
light unit$,$ for  $U^m$ and use a $L_{sc}^s$ for $U^s.$ 
It appears for this particular scenario$,$ that $L_{sc}^s$ may be considered 
the private unit of length in the NSP-world, such as $L,$  used to measure NSP-world 
light-path length. 
The  ``wavelength'' $\lambda$ 
of any 
light source must also 
be measured in the same light units. Let $\lambda = N^sL_{sc}^s.$  
Taking into consideration a unit 
conversion 
factor $k$ between the unknown NSP-world private units$,$ such that $\st {k
L_{sc}^s }= U^s$, the number of light waves
in $s$-laboratory would be $A^s \st {k L_{sc}^s}/N^s\st {kL_{sc}^s} = 
A^s/N^s,$ 
where $A^s$ is a pure number such that $A^s\st {kL_{sc}^s}$ is the ``path-length'' using the 
units in the $s$-system. In the moving system$,$ assuming that this 
simple aspect of light propagation holds in the NSP-world and the 
N-world which we did to obtain the derivation in section 3$,$ 
it is claimed that substitution yields  
$\st {\beta^{-1}A^skL_{sc}^s}/\st{\beta^{-1}N^s kL_{sc}^s}= A^s\st {\beta^{-1}kL_{sc}^s}/N^s\st{{\beta}^{-1}kL_{sc}^s}= A^s/N^s.$ 
Thus there would be 
no difference in the number of light waves in any case where the 
experimental set up 
involved the sum of light paths each of which corresponds to the to-and-fro 
process [1: 24]. Further$,$ the same conclusions would be reached using (8.2). 
not 
relevant to a Sagnac type of experiment. However$,$ this does not mean that a
similar derivation involving a polygonal propagation path cannot be 
obtained. [Indeed$,$ this may be a consequence of a result to be derived in 
article 3. However$,$ see note 8 part 4$,$ p. 80.] \par  
Where is the logical error in the above argument?
The error is the object upon which the $\st {\beta}^{-1}$ operates. Specifically
(6.13) states that \pars

$$ \st {\beta}^{-1}(A^skL_{sc}^s)\ \ \ {\buildrel ({\rm emis}) \over \longleftrightarrow}
\ \ \ \beta^{-1}(L\Pi^{(s)}) = (\beta^{-1}L)\Pi^{(s)} \ {\rm and}\eqno (8.3)$$ 
$$\st {\beta}^{-1}(N^skL_{sc}^s)\ \ \ {\buildrel ({\rm emis}) \over \longleftrightarrow}
\ \ \ \beta^{-1}(L\Pi_1^{(s)})= (\beta^{-1}L)\Pi_1^{(s)}.\eqno (8.4)$$ 
\noindent It is now rather obvious that the two (emis) aspects of the M-M experiment 
nullify each other.  
 Also for no finite  
$w$ can $\beta \approx 0.$ There is a great difference between the 
propagation properties in the NSP-world and the N-world. For example$,$ the 
classical Doppler effect is an N-world effect relative to linear propagation. 
{\bf Rather than indicating that the NSPPM is not 
present$,$ the M-M results indicate indirectly that the NSP-world NSPPM 
exists.}\par 

Apparently$,$ the well-known Ives-Stillwell$,$ and all similar$,$ experiments 
used in an attempt to verify such things as the relativistic redshift are of 
such a 
nature that they 
eliminate other effects that motion is assumed to have  upon the scenario 
associated electromagnetic 
{\it propagation.}  What was shown is that 
the frequency $\nu$ of the canal rays vary with respect to a representation for 
$v_E$ measured from electromagnetic theory in the form 
$\nu_m=\st {\beta}^{-1} \nu_s.$ First$,$ we must investigate what the so-called time 
dilation statement (8.2) means. What it means is exemplified by (6.14) and 
how the human mind comprehends the measure of  ``time.'' 
In the scenario associated (8.2) expression$,$ 
for the right and left-sides to be comprehensible$,$
the expression  should be 
conceived of as  a measure that originates with infinitesimal light-clock behavior.
It is the 
experience with a specific unit and the number of them that ``passes'' that 
yields the intuitive concept of ``observer time.'' On the other hand$,$ for some purposes 
or as some authors assume$,$ (8.2) might be 
viewed as a change in a time unit $T^s$ rather than in an infinitesimal light-clock. 
Both of these interpretations can be incorporated into a frequency statement.
First$,$ relative to the frequency of light-clock counts$,$
for a fixed stationary unit of time 
$T^s,$ (8.2) reads 
$$\st {\beta}^{-1}C_{sc}^s/T^s\doteq C^m_{sc}/T^s \Rightarrow\st {\beta}^{-1}
C_{sc}^s\doteq C^m_{sc}.\eqno (8.5)$$ \par

\underbar{But} according to (6.18)$,$ the $C_{sc}^s$ and $C^m_{sc}$ correspond to 
infinitesimal light-clocks measures and nothing more 
than that. Indeed$,$ (8.5) has nothing to do with  the concept of absolute ``time''
only with the different infinitesimal light-clocks that need to be used due to relative motion. 
This requirement may be due to (emis). Indeed$,$ the ``length contraction'' 
expression (8.1) and the  ``time dilation'' expression (8.2) have nothing 
to do with either absolute length or absolute time. These two expressions are both saying the 
same thing from two different viewpoints. There is an alteration due to the 
(emis). [Note that the  second $\doteq$ in (8.5) depends upon the $T^s$ 
chosen.] \par  
On the other hand$,$ for a relativistic redshift type experiment$,$ the usual interpretation is that 
$\nu_s\doteq p/T^s$ 
and $\nu_m\doteq  p/T^m.$ This leads to $p/T^m \doteq \st {\beta}^{-1} 
p/T^s\Rightarrow T^m  
\doteq \st {\beta}T^s.$   
Assuming that all frequency alterations due 
to (emis) have been eliminated then this is interpreted to mean that  
``time'' is slower in the moving excited hydrogen atom 
 than in the ``stationary'' laboratory. When compared to (8.5)$,$ 
there is the ambiguous interpretation in that the $p$ is considered the same for both sides
(i.e. the 
concept of the frequency is not altered by NSPPM motion). It is 
consistent with all that has come before that the Ives-Stillwell result be 
written as $\nu_s\doteq p/T^s$ 
and that $\nu_m\doteq q/T^s,$ where ``time'' as a general notion is not altered. This leads to the expression 
$$\st {\beta}^{-1}p \doteq q\  [=\ {\rm in\ the\ limit}].\eqno (8.6)$$\par 
Expression (8.6) does not correspond to a concept of  ``time'' but rather to the 
concept of alterations in emitted frequency due to (emis).   
One$,$ therefore$,$ has an ambiguous  
interpretation that in an Ives-Stillwell scenario
the number that represents the frequency of light emitted from an atomic unit moving with velocity 
$\omega$ with respect to the NSPPM is altered due to (emis).  
This (emis) alteration depends upon $K^{(3)}.$ It is critical that the two 
different infinitesimal light-clock interpretations be understood. One 
interpretation is relative to electromagnetic {\it propagation} theory. In this case$,$ the light-clock concept is taken in its most literal form. The 
second interpretation is relative to an infinitesimal light-clock as an {\it analogue} 
model. 
This means that the cause need not be related to propagation but is more 
probably due to how individual constituents interact with the NSPPM. 
The 
exact nature of this interaction and a non-ambiguous approach needs further investigation based upon 
constituent models since the analogue model specifically denies that there is 
some type of 
 {\it absolute time} dilation but$,$ rather$,$ signifies the existences of other 
possible causes. [In article 3$,$ the  $\nu_m=\st {\beta}^{-1} \nu_s$ 
is formally and non-ambiguously derived from a special line-element$,$ a universal 
functional requirement and Schr\"odinger's equation.] \par 

In our analysis it has been 
assumed that $F_1$ is stationary in the NSP-world NSPPM. It is clear$,$ 
however$,$ that under our assumption that the scalar velocities in the NSP-world
are additive with respect to linear motion$,$ then if $F_1$ has a velocity
$\omega$ with respect to the NSPPM and  $F_2$ has the velocity 
$\omega^\prime,$ then 
it follows that the light-clock counts for $F_1$ require the use of a different light-clock with respect to a stationary $F_0$ due to the (emis) and the light-clocks for 
$F_2$ have been similarly changed with respect to a stationary $F_0$ 
due to (emis). Consequently$,$ a light-clock related expressed by $K^{(3)}$ is 
the result of the combination$,$ so to speak$,$ of these two (emis) influences. The relative NSPPM velocity $\omega_2$ of $F_1$ with respect to $F_2$ which yields the difference between these influences is that which would satisfies the additive rule for three linear positions.  
\par
 As previously stated$,$ within the NSP-world
relative to electromagnetic propagation, observer scalar velocities are either 
additive or related as discussed above. Within the N-world$,$ this last 
statement need not be so. Velocities of individual entities are modeled 
by either 
vectors or$,$ at the least$,$ by signed numbers. Once the N-world expression is 
developed$,$ then it can be modified in accordance with the usual (emis) 
alterations$,$ in which case the velocity statements are N-world Einstein 
measures. For example$,$ deriving the so-called relativistic Dopplertarian 
effect$,$
the combination of the classical and the relativistic redshift$,$
by means of a NSPPM argument such as appears in [7] where it is 
assumed that the light propagation laws with respect to the photon concept
in the NSP-world
are the same as those in the N-world$,$ is in logical error. Deriving the 
classical Doppler effect expression then$,$ when physically justified$,$ 
making the wave number alteration
in accordance with the (emis) would be the correct logic needed to obtain 
the relativistic Dopplertarian effect. [See note 6.]\par
Although I will not$,$ as yet$,$  re-interpreted Special Relativity results with 
respect to this purely electromagnetic interpretation$,$ it is interesting to 
note the following two re-interpretations. The so-called variation of  
``mass'' was$,$ in truth$,$  originally derived for imponderable matter 
(i.e. elementary 
matter.) This would lead one to believe that the so-called rest mass and its 
alteration$,$ if experimentally verified$,$ is really 
a manifestation of the electromagnetic nature of such elementary matter. 
Once again the 
so-called mass alteration can be associated with an (emis)
concept. 
The $\mu$-meson decay rate may also show the same type of alteration as 
appears to be the case in an Ives-Stillwell experiment.   It does not take 
a great 
stretch of the imagination to again attribute the apparent
alteration in this rate to an (emis) process. This would lead to 
the possibility that such decay is controlled by electromagnetic properties. 
Indeed$,$ in order to conserve various things$,$  $\mu$-meson decay
is said to lead to the 
generation of the neutrino and antineutrino. 
[After this paper was completed$,$ a method was discovered that establishes that 
predicted mass and decay time alterations are (emis) effects. The derivations 
are found in article 3.]\par
I note that such things as neutrinos and antineutrinos need not exist. Indeed$,$ 
the nonconservation of certain quantities for such a scenario leads to the
conclusion that subparticles exist within the NSP-world and carry off 
the ``missing'' quantities. Thus the invention of such objects may 
definitely be considered as only a bookkeeping technique.\par
As pointed out$,$ all such experimental verification of the properly interpreted 
transformation equations can be considered as indirect evidence that the 
NSP-world NSPPM exists. But none of these results should be extended 
beyond the experimental scenarios concerned. Furthermore$,$ I conjecture that 
no matter how the human mind attempts to explain the (emis) in terms of a 
human language$,$ it will always be necessary to postulate some interaction
process with the NSPPM without being able to specifically describe this 
interaction in terms of more fundamental concepts. Finally$,$ the MA-model
specifically states that the Special Theory is a local theory and should not
be extended, without careful consideration, beyond a local time interval $[a,b].$\pars
\leftline{\bf 9. Reciprocal Relations}\pars
As is common to many mathematical models$,$ not all relations generated by 
the mathematics need to correspond to physical reality. This is the modern
approach to the length contradiction controversy [6]. Since this is a 
mathematical model$,$ there is a theory of correspondence between the physical 
language and the mathematical structure. This correspondence should be 
retained throughout any derivation. This is a NSPPM theory and what 
is stationary or what is not stationary with 
respect to the NSPPM must be maintained throughout any correspondence.
This applies to such reciprocal relations as $$\st {\beta}^{-1}({\overline{t}}^{(m)}_E -t^{(m)}_E)= 
\overline{t}^{(s)} - 
t^{(s)} \eqno (9.1)$$ 
and
$$\st {\beta}^{-1}({\overline{x}}^{(m)}_E -x^{(m)}_E)= 
\overline{x}^{(s)} - 
x^{(s)} \eqno (9.2)$$     
  Statement (8.1) and (9.1) [resp. (8.2) and (9.2)] both hold from the 
NSPPM viewpoint only when 
$v_E = 0$ since it is not the question of the N-world viewpoint of relative 
velocity but rather the viewpoint that $F_1$ is fixed and $F_2$ is not fixed 
in the 
NSPPM or $\omega \leq \omega^\prime.$ The physical concept of the $(s)$ 
and $(m)$ must be maintained throughout the physical correspondence. 
 Which expression would hold for a particular scenario depends upon 
laboratory confirmation. This is a scenario associated theory.  
All of the laboratory scenarios discussed in this paper use infinitesimalized (9.1) and (9.2) as derived from line-elements and the ``view'' or comparison is always made relative to the $(s)$. Other authors$,$ such as Dingle [1] and Builder [7]$,$ have, in a absolute sense, excepted one of these sets of equations, without derivation, rather the other set. I have not taken this  
stance in this paper.\pars

One of the basic controversies associated with the Special Theory is whether 
(8.2) or (8.1) [resp. (9.1) or (9.2)] actually have physical meaning. The 
notion is that either ``length'' is a fundamental concept and ``time'' is 
defined in terms of it$,$ or ``time'' is a fundamental concept and length is 
defined in terms of it. Ives$,$ and many others assumed that the 
fundamental notion is length contraction and not time dilation. The modern 
approach is the opposite of this. Length contraction in the N-world has no physical meaning$,$ but time dilation does [6].  
We know that time is often defined in terms of 
length and velocities. But$,$ the length or time being considered here is 
Einstein length or Einstein time. This is never mentioned when this problem is 
being considered. As discussed at the end of section 3$,$ Einstein length is 
actually defined in terms of infinitesimal light-clocks or in terms of 
the Einstein velocity and Einstein time. As shown after equation (8.2) is considered, it is only infinitesimal light-clock ``time'' that is altered and length altertions is but a technical artefact.   The changes in the infinitesimal light-clock  counts yields an analogue model for physical changes that cause Special 
Theory effects.  [See note 7.] \parm

$\{$Remark: Karl Popper notwithstanding$,$ it is not the sole purpose of 
mathematical models to predict natural system behavior. The major purpose is to 
maintain logical rigor and$,$ hopefully$,$ \underbar{when applicable} to discover 
new properties for natural systems. I have used in this speculation a 
correspondence theory that takes the stance that any verifiable Special Theory
effect is electromagnetic in character rather than a problem in measure. 
However$,$ whether such effects are simply effects relative to the
propagation of electromagnetic information or whether they are effects
relative to the constituents  involved cannot be directly obtain from the 
Special Theory. All mathematically stated effects involve
the Einstein measure of relative velocity$,$ $v_E$ -- a propagation related 
measure. The measure of an effect should also be done in accordance with 
electromagnetic theory.   
As demonstrated$,$ the Special Theory should not 
be unnecessarily applied to the behavior of all nature systems since it is 
related to electromagnetic interaction; unless$,$ of course$,$ all natural systems 
are electromagnetic in character. 
Without strong justification$,$ the assumption that one theory does 
apply to all scenarios is one of the 
greatest errors in mathematically modeling.    
But$,$ if laboratory experiments verify 
that alterations are  taking place in measured quantities and these 
variations are 
\underbar{approximated} in accordance with the Special Theory$,$
then this would indicate that either the alterations are related to 
electromagnetic propagation properties or the constituents  have an 
appropriate electromagnetic character.$\}$
\parm

\centerline{NOTES}
\medskip
[1] (a) Equation (3.9) is obtained as follows: since $t \in [a,b],$  
$t$ finite and not infinitesimal. Thus  division by $t$ preserves $\approx.$
Hence$,$
$$\left[t\left({{\hyper s(t + dt) - s(t)}\over {dt}}\right)- s(t)\right]/{t^2} 
\approx {{\ell(t)}\over{t}}.\eqno (1)$$ Since $t$ is an arbitrary standard number 
and $dt$ is 
assume to be an arbitrary and appropriate nonzero infinitesimal and 
the function $s(t)/t$ is differentiable$,$ the standard part of the 
left-side equals 
the standard part of the right-side.  Thus 
$${{d(s(t)/t)}\over {dt}} = {{v(t)}\over {t}}, \eqno (2)$$ 
for each $t \in [a,b].$ 
By *-transfer$,$ equation (3.9) holds for each $t \in \Hyper [a,b].$\par
(b) Equation 
(3.10) is then obtained by use of the *-integral and the fundamental theorem 
of integral calculus *-transferred to the NSP-world. It is useful to 
view the definite integral over a standard interval say $[t_1,t]$ as an 
operator$,$ at least$,$ defined on the set $C([t_1,t], \real )$ of all continuous real valued 
functions 
defined on $[t_1,t].$ Thus$,$ in general$,$ the fundamental theorem of integral 
calculus can be viewed as the statement that $(f^\prime , f(t) - f(t_1)) \in 
\int_{t_1}^t.$ Hence $\Hyper (f^\prime , f(t) - f(t_1)) \in 
\Hyper \int_{t_1}^t \Rightarrow (\hyper f^\prime , \Hyper (f(t) -
f(t_1))) \in \hyper \int_{t_1}^t \Rightarrow   (\hyper f^\prime , f(t) -
f(t_1)) \in \hyper \int_{t_1}^t.$\pars
(c)  To obtain the expressions in (3.19)$,$
consider $f(x) = 1/x.$ Then $\hyper f$  is limited and S-continuous on 
$\Hyper [a,b].$ Hence 
$(\hyper f, \ln t_2 - \ln t_1) \in \hyper \int_{t_1}^{t_2}.$  Hence
$\st {(\hyper f, \ln t_2 - \ln t_1)} = (f, \ln t_2 - \ln t_1) \in 
\int_{t_1}^{t_2}.$ Further (3.19) can be interpreted as an interaction property. \pars
[1.5] Infinitesimal light-clocks are based upon the QED model as to how electrons are kept in a range of distances in a hydrogen atom proton. The back-and-forth exchanges of photons between a proton and electron replaces ``reflection'' and the average distance between the proton and electron is infinitesimalized to the $L$. In this case, the proton and electron are also infinitesimalized. The large number of such interchanges over a second, in the model, is motivation for the use of the members of $\nat^+_\infty$ as count numbers.\pars

[2] The basic theorem that allows for the entire concept of infinitesimal 
light-clocks and the analysis that appears in this monograph has not been 
stated. As taken from ``The Theory of Ultralogics,''the theorem, for this application, is:\par 
{\bf Theorem 11.1.1} {\sl Let $10^\omega \in \nat_\infty.$ Then for each $ r \in \real$ there exists an $x \in \{2m/10^\omega \mid(2m \in \Hyper {\b Z}) \land
(\vert 2m\vert < \lambda 10^\omega)\},$ for any $\lambda \in \nat_\infty,$ such that $x \approx r$ (i.e. $ x \in \monad r.)$}\pars
Theorem 11.1.1 holds for other members of $\nat_\infty.$ Let $L =1/10^\omega$ where $\omega$ is any hyperreal infinite natural number 
(i.e. $\omega \in \nat_\infty).$ Hence$,$ by this theorem$,$ for any positive real 
number $r$ there exists some $m \in \nat_\infty$ such that $2\st {m/10^\omega} = r.$
I point out that for this nonzero case it is necessary that $m \in 
\nat_\infty$ for if  $m \in \nat,$ then $\st{m/10^\omega} = 0.$ Since $c =\st {L/u},$ then $2\st {um} = 2\st {(L/c)m} = t = r/c$ as required. Thus, the infinitesimal light-clock determined length $r$ and interval of time $t$ are determined by the difference in infinitesimal light-clock counts $2m = (\lambda -\eta)$. Note that our approach allows the calculus to model this behavior by simply assuming that the standard functions are differentiable etc. \pars 

[2.5] (4 JUN 2000) Equating these counts here and elsewhere is done so that the ``light pulse'' is considered to have a ``single instantaneous effect'' from a global viewpoint and as such is not a signal in that globally it contains no information. 
Thus additional analysis is needed before one can state that the Special Theory applies to informational transmissions. It's obvious from section 7 that the actual value for $c$ may depend upon the physical application of this theory.\pars
[3] At this point and on$,$ the subscripts 
on the $\tau$ have a different meaning than previously indicated. The 
subscripts denote process numbers while the superscript denotes the position 
numbers. For example$,$ $\tau^2_{12}$ means the light-clock count number when the second 
light pulse leaves
$F_2$ and $\tau^2_{31}$ would mean the light-clock count number when the first 
light pulse returns to position $F_2.$\par
  The additional piece of each subscript
denoted by the $a$ on this and the following pages indicates$,$ what I thought 
was obvious from the lines that follow their introduction$,$ that these are 
approximating numbers that are 
infinitesimally near to standard NSP-world number obtained by taking the 
standard part. \pars
[4] Note that such infinite hyperreal numbers as $\Pi^{(2)}_3$ (here and 
elsewhere) denote the 
difference between two infinitesimal light-clock counts and since we are 
excluding the finite number infinitesimally near to 0$,$ these numbers must be 
infinite hyperreal. Infinitesimal light-clocks can be assumed to measure this 
number by use of a differential counter. BUT it is always to be conceived of 
as an infinitesimal light-clock  ``interval'' (increment$,$ difference$,$ etc.) It 
is important to recall this when the various line-elements in the next 
article are considered. \pars
[5] This result is 
obtained as follows: since $t_a \leq t_2 \leq t_b,$  it follows that 
$\vert t_a - t_2 \vert < \delta,\ \vert t_b - t_2 \vert 
< \delta.$ Hence by *-transfer, $\vert \Hyper v(t_2) - \Hyper v(t_a) \vert < 
\eps/3,\ \vert \Hyper v_1(t_b) - \Hyper v_1(t_2) \vert < 
\eps/3.$ Since we assume arbitrary $\eps/3$ is a standard positive number$,$ 
then $\Hyper v(t_a) = 
\ell(t_a) \approx \Hyper v_1(t_b) = \ell_1(t_b) \Rightarrow 
\vert \Hyper v(t_a) - \Hyper v_1(t_b) \vert< \eps/3.$ Hence $\vert \Hyper 
v(t_2) - \Hyper v_1(t_2)\vert < \eps.$                                                           
 \pars

[6] In this article$,$ I mention that all previous derivations for 
the complete Dopplertarian effect (the N-world and the transverse) are in 
logical error. Although there are various reasons for a redshift not just
the Dopplertarian$,$ the electromagnetic redshift based solely upon properties 
of the NSPPM can be derived as follows: \par
 (i) let $\nu^s$ denote the  ``standard'' laboratory frequency for radiation 
emitted from an atomic system. This is usually determined by the observer. 
The  
NSP-world alteration in emitted frequency at an atomic structure 
due to (emis) is  $\gamma\nu^s = \nu^{\rm radiation},$ where $\gamma = 
\sqrt {1-v_E^2/c^2}$ and $v_E$ is the Einstein measure of the relative velocity 
using light-clocks only. \pars
(ii) Assuming that an observer is observing this emitted radiation in a 
direct line with the propagation and the atomic structure is receding with 
velocity $v$ from the 
observer$,$ the frequency of the electromagnetic 
propagation$,$ within the 
N-world$,$ is altered compared to the observers standards. This alteration is $\nu^{\rm radiation}
(1/(1+v/c)) = \nu^{\rm received}.$ Consequently$,$ this yields the total 
alteration as $\gamma\nu^s(1/(1+v/c)) = \nu^{\rm received}.$ Note that $v$ 
is measured in the N-world and can be considered a directed velocity. 
Usually$,$ if due to 
the fact that we are dealing with electromagnetic radiation$,$  
we 
consider $v$ the Einstein measure of linear velocity (i.e. $v = v_E$)$,$ 
then the total Dopplertarian effect for $v \geq 0$ can be written as
$$\nu^s\left({{1-v_E/c}\over{1 + v_E/c}}\right)^{1/2} = \nu^{\rm 
received}.\eqno (3)$$\par
It should always be remembered that there are other reasons$,$ such as the 
gravitational redshift and others yet to be analyzed$,$ that can mask 
this total Dopplertarian redshift. \pars

[7]  A question that has been asked relative to the new 
derivation that yields Special Theory resilts is why in the 
N-world do we have the apparent nonballistic effects associated with 
electromagnetic radiation? In the derivation$,$ the opposite was assumed for the 
NSP-world
monadic clusters. The constancy of the {\it measure}$,$ by light-clocks and the 
like$,$ of the $F_1 \to F_2,\ F_2 \to F_1$ velocity of electromagnetic radiation was modeled by 
letting $\st {t_a} = \st {t_b}.$ As mentioned in the section on the Special 
Theory$,$  the Einstein velocity measure transformation expression can be 
obtained prior to embedding the world into a hyperbolic velocity space. It is 
obtained by considering three in-line standard positions $F_1,\ F_2, \ F_3$ 
that have the NSP-world velocities $w_1$ for $F_2$ relative to $F_1$$,$
$w_2$ for $F_3$ relative to $F_2$ and the simple composition $w_3= w_1 + w_2$
for $F_3$ relative to $F_1$. Then simple substitution in this expression
yields 
$$v_E^{(3)}= (v_E^{(1)}+v_E^{(2)})/\left(1 + 
{{v_E^{(1)}v_E^{(2)}}\over{c^2}}\right).\eqno (4)$$
 This relation is telling us 
something about the required behavior in the N-world of electromagnetic 
radiation. To see that within the N-world we need to assume   for 
electromagnetic radiation effects the nonballistic property$,$  simply  
let $v_E^{(2)}=c$ or $ v_E^{(2)}\doteq c.$ Then $v_E^{(3)}=c,$ or 
$\doteq c.$ Of course$,$ the reason we do not have a contradiction is that 
we have two distinctly different views of the behavior of electromagnetic 
radiation$,$ the NSP-world view and the N-world view. Further$,$ note how$,$ for 
consistency$,$ the velocity of electromagnetic radiation is to be measured. It 
is measured by the Einstein method$,$ or equivalent$,$ relative to a 
to-and-fro path and measures of ``time'' and ``distance'' by means of a 
(infinitesimal) light-clock counts. Since one has the NSPPM, then letting $F_1$ be fixed in that medium, assuming that ``absolute'' physical standards are measured at $F_1,$ equation (4) indicates why, in comparison, physical behavior varies at $F_2$ and $F_3$. The hyperbolic velocity space properties are the cause for such behavior differences.  \pars
  
I am convinced that the dual character of the Special 
theory derivation requires individual reflection in order to be understood fully. In the NSP-world$,$ 
electromagnetic radiation behaves in one respect$,$ at least$,$ like a particle in 
that it satisfies the ballistic nature of particle motion.  The reason that 
equation (3) is derivable is due to the definition of Einstein time. But 
{\it Einstein time$,$ as measured by electromagnetic pulses$,$ models the 
nonballistic 
or one and only one wave-like property in that a wave front does not partake of the 
velocity of the source.} This is the reason why  
I wrote that {\it a NSPPM disturbance would trace the 
same operational linear light-clock distance.} The measuring light-clocks are 
in the N-world in this case. $F_1$ is modeled as fixed in the NSPPM and 
$F_2$ has an NSP-world relative velocity. The instant the light pulse is 
reflected 
back to $F_1$ it does not$,$ from the N-world viewpoint$,$ partake of the N-world
relative velocity and therefore traces out the exact same apparent N-world 
linear path. The position $F_2$ acts like a virtual position having no other 
N-world effect upon the light pulse except a reversal of direction. \pars
[8] This expression implies that the ``$c$'' that appears here and elsewhere is to be measured by infinitesimal light-clocks.  As noted $u \approx L/c,$ but infinitesimal light-clock construction yields that $u =L/c.$ For a fixed $L$, from the NSPPM  viewpoint, $u$ is fixed. Notice that $t^{(i)}\approx u(2\eta^{(i)}) = u(\gamma^{(i)}),\ \gamma^{(i)}\in \nat^+_\infty.$ \parm
\centerline{\bf REFERENCES}
\parm
\noindent {\bf 1} Dingle$,$ H.$,$ {\it The Special Theory of Relativity}$,$ Methuen's 
Monographs on Physical Subjects$,$ Methuen \& Co. LTD$,$ London$,$ 1950.\pars
\noindent {\bf 2} Herrmann$,$ R. A.$,$ {\it Some Applications of Nonstandard 
Analysis to Undergraduate Mathematics -- Infinitesimal Modeling}$,$ 
Instructional Development Project$,$ Mathematics Dept. U. S. Naval Academy$,$ 
Annapolis$,$ MD 21402-5002$,$ 1991. http://arxiv.org/abs/math/0312432 \pars
\noindent {\bf 3} Herrmann$,$ R. A.$,$ Fractals and ultrasmooth microeffects. {\it 
J. Math. Physics.}$,$ {\bf 30}(1989)$,$ 805--808.\pars
\noindent {\bf 4} Herrmann$,$ R. A.$,$  Physics is legislated by a cosmogony. {\it
Speculat. Sci. Technol.}$,$ {\bf 11}(1988)$,$ 17--24.\pars
\noindent {\bf 5} Herrmann$,$ R. A.$,$ Rigorous infinitesimal modelling$,$ {\it Math. 
Japonica} 26(4)(1981)$,$ 461--465.\pars
\noindent {\bf 6} Lawden$,$ D. F.$,$ {\it An Introduction to Tensor Calculus$,$ 
Relativity and Cosmology}$,$ John Wiley \& Son$,$ New York$,$ 1982.\pars
\noindent {\bf 7} Prokhovnik$,$ S. J.$,$ {\it The Logic of Special Relativity}$,$ 
Cambridge University Press$,$ Cambridge$,$ 1967.\pars
\noindent {\bf 8} Stroyan$,$ K. D. and Luxemburg$,$ W. A. J. 1976. {\it Introduction to 
the Theory of Infinitesimals}$,$ Academic Press$,$ New York.\pars
\noindent {\bf 9} Herrmann$,$ R. A.$,$ {\it The Theory of Ultralogics}$,$ (1992)
 \hfil\break
http://arxiv.org/abs/math.GM/9903081\hfil\break
http://arxiv.org/abs/math.GM/9903082\par
\pars
\noindent {\bf 10} Davis M.$,$ {\it Applied Nonstandard Analysis$,$} John Wiley 
\& Sons$,$ New York$,$ 1977.
\medskip

\centerline{\bf Appendix-A}\parm
\noindent {\bf 1. The Need for Hyperbolic Geometry}\parm

In this appendix$,$ it is shown that from equations (3.21) and (3.22) the Lorentz transformation are derivable. All of the properties for the Special Theory are based upon ``light'' propagation. In Article 2$,$ the concern is with two positions $F_1,\ F_2$ in the NSPPM within the NSP-world and how the proposed NSPPM influences such behavior. Prior to applications to the N-world$,$ with the necessity for the N-world Einstein measures$,$ the NSPPM exhibits infinitesimal behavior and special NSPPM non-classical global behavior. The behavior at specific moments of NSPPM time for global positions and classical uniform velocities are investigated. \par 

The following is a classical description for photon behavior. Only NSPPM relative velocities (speeds) are being considered. 
Below is a global diagram for four points that began as the corners of a square, where $u$ and $\omega$ denote uniform relative velocities between point locations and no other point velocities are considered. The meanings for the symbolized entities are discussed below. \par\bigskip

\line{\hskip 1.35in$\bullet F_1\ t$ $\sim\kern -0.25em\sim\kern -0.25em\longrightarrow $\hskip 1.25in$\omega \longrightarrow \bullet F_2(t(p_1))\sim\kern -0.25em\sim\kern -0.25em\longrightarrow$\hfil}
\vskip 0.35in 
\line{\hskip 1.35in $\bullet F_1'\ t'\sim\kern -0.25em\sim\kern -0.25em$\hfil}
\line{\hskip 1.20in $u \ \downarrow$\hfil}
\line{\hskip 1.35in $\bullet F_1'\ t(p_2)$\hskip 2.00in $\omega \longrightarrow \bullet F_2'(t(p_2))$\hfil} 

\line{\hskip 1.20in $u \ \downarrow$ \hskip 2.76in$\ \downarrow\ u$\hfil}\par\bigskip

Consider the following sequence of (conceptual) NSPPM time-ordered events. First, the N-world position points $F_1,\ F_2,\ F_1',\ F_2'$ are stationary with respect to each other and form the corners of a very small rhombus, say the side-length is the average distance $d$ between the electron and proton within an hydrogen atom. The sides are $\overline{F_1,F_2}, \ \overline{F_2F_2'},\ \overline{F_2',F_1'},\ \overline{F_1'F_1}$. At the NSPPM time $t_g$, the almost coinciding $F_2,F_2'$ uniformally recede from the almost coinciding $F_1,F_1'$ with constant velocity $\omega$.  At a time $t > t_g,$ where the distance between the two groups is significantly greater than $d$, one process occurs simultaneously. The point $F_1'$ separates from $F_1$ with relative velocity $u$ and $F_2'$ separates from $F_2$ with a relative velocity $u$. [Using NSP-world processes, such simultaneity is possible relative to a non-photon transmission of information (Herrmann, 1999).] At any time $\geq t,$ the elongating line segments $\overline{F_1F_1'}$ and $\overline{F_2F_2'}$ are parallel and they are not parallel to the parallel elongating line segments $\overline{F_1F_2}$ and $\overline{F_1'F_2'}.$ \par

At NSPPM time $t,$ a photon $p_1$ is emitted from $F_1$ towards $F_2$ and passes through $F_2$ and continues on. As $F_1'$ recedes from $F_1$, at $t' >t,$ a photon $p_2$ is emitted from $F_1'$ towards $F_2'$. The original classical photon-particle property that within a monadic cluster photons prorogate with velocity $\omega +c$ is extended to this global environment. [Again there are NSP-world processes that can ensure that the emitted photons acquire this prorogation velocity (Herrmann, 1999).] Also, this classical photon-particle property is applied to $u$. Thus, photon $p_2$ is assumed to take on an additional velocity component $u$. Photon, $p_1$, passes through $F_2$ at the NSPPM time $t(p_1)$. Then $p_2$ is received at point $F_2'$ at time $t(p_2)$.  \par

Classically, $t(p_1') > t(p_1).$ From a viewpoint relative to elongating $\overline {F_1F_1'}$, the distance between the two photon-paths of motion measured parallel to elongating $\overline {F_1F_1'}$  is $u(t(p_2) -t).$ On the other hand, from the viewpoint of elongating $\overline{F_1F_1'}$, the distance between photon-paths, if they were parallel, is $u(t' - t).$ 
By the relativity principle, from the viewpoint of $F_1'$, the first equation in (3.19) should apply. Integrating, where $\st {\hyper {v(t_a))}} = c$, one obtains $u(t(p_2)- t)) = ue^{\omega/c}(t' - t).$ [Note: No reflection is required for this restricted application of (3.19).] This result is not the classical expression $u(t(p_2)- t)).$  For better comprehension, use infinitesimal light-clocks to measure NSPPM time. Then using the same NSPPM process that yields information instantaneously throughout the standard portion of the NSPPM, all clocks used to determine these times can be set at zero when they indicate the time $t$. This yields that the two expressions for the distance are $ut(p_2)$ and $ue^{\omega/c}t'.$ However, the classical expression $ut(p_2)$ has the time $t(p_2)$ dependent upon both $\omega$ and, after the $t'$ moment, upon $u.$ But, for the relativistic expression, the $t'$ is neither dependent upon the $u$ velocity after $t'$ nor the $\omega$ and the factor $e^{\omega/c}$ has only one variable $\omega.$ What property does this NSPPM behavior have that differentiates it from the classical?\par

Consider the two velocities $u$ and $ue^{\omega/c}.$ These two velocities only correspond when  $\omega =0.$ Hence, if we draw a velocity diagram, one would conclude that, in this case, the velocities are trivially ``parallel.''  Using Lobatchewskian's horocycle construction$,$ Kulczycki (1961) shows that for ``parallel  geometric'' lines in hyperbolic space$,$ the distance between each pair of such lines increases (or decreases) by a factor $e^{x/k},$ as one moves an ordinary distance $x$ along the lines and $k$ is some constant related to the $x$ unit of measurement. Phrasing this in terms of velocities, where $x = \omega$ and $k = c$, then, for this case, the velocities, as represented in the NSPPM by standard real numbers, appear to satisfy the properties for an hyperbolic velocity-space. Such velocity behavior would lead to this non-classical NSPPM behavior.\par 

When simple classical physics is applied to this simple Euclidian configuration within the NSPPM, then there is a transformation $\rm \Phi\colon NSPPM \to$ N-world, which is characterized by hyperbolic velocity-space properties. This is also the case for relative velocity and collinear points, which are exponentially related to the Einstein measure of relative velocity in the N-world. In what follows, this same example is used but generalized slightly by letting $F_1$ and $F_2$ coincide. \parm

\noindent {\bf 2. The Lorentz Transformations}\parm
Previously, we obtained the expression that $t_2 = \sqrt {t_1t_3}.$ The Einstein measures are defined formally as 
$$\cases{t_E = (1/2)(t_3 + t_1)&\cr
         r_E = (1/2)c(t_3 - t_1)&\cr
         v_E = r_E/t_E,\ {\rm where\ defined.}&\cr}\eqno (A1)$$
Notice that when $r_E = 0,$ then $v_E = 0$ and $t_E = t_3 = t_1= t_2$ is not Einstein measure.\par 

\noindent The Einstein time $t_E$ is obtained by considering the ``flight-time'' that would result from using one and only one wave-like property not part of the NSPPM but within the N-world. This property is that the $c$ is not altered by the velocity of the source. This Einstein approach assumes that the light pulse path-length from $F_1$ to $F_2$ equals that from $F_2$ back to $F_1.$ Thus, the Einstein flight-time used for the distance $r_E$ is $(t_3 - t_1)/2$. The $t_E,$ the Einstein time corresponding to an infinitesimal light-clock at $F_2,$ satisfies $t_3 - t_E = t_E - t_1.$ From $(A1),$ we have that 
$$t_3 = (1+v_E/c)t_E\ {\rm and } \ t_1 = (1 - v_E/c)t_E, \eqno (A2)$$ and$,$ hence$,$ $t_2 = (\sqrt {1 - v_E^2/c^2})t_E.$ Since $e^{\omega/c} = \sqrt {t_3/t_1},$ this yields
$$e^{\omega/c} = \left({{1 + v_E/c}\over{1 - v_E/c}}\right)^{(1/2)}. \eqno (A3)$$ \pars

Although it would not be difficult to present all that comes next in terms of the nonstandard notions$,$ it is not necessary since all of the functions being consider are continuous and standard functions. The effect the NSPPM has upon the N-world are standard effects produced by application of the standard part operator ``st.''\par

From the previous diagram, let $F_1$ and $F_2$ coincide and not separate. Call this location $P$ and consider the diagram below. This is a three position classical NSPPM light-path and relative velocity diagram used for the infinitesimal light-clock analysis in section 6 of Article 2. This diagram is not a vector composition diagram but rather represents linear light-paths with respect to Einstein measures for relative velocities. It is also a relative velocity diagram to which hyperbolic ``geometry'' is applied. \par\bigskip
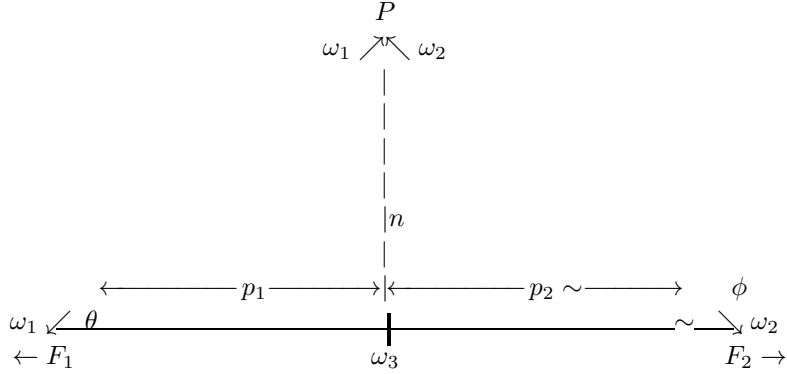
\line{\hfil$P$\hfil} 
\line{\hfil $\omega_1\nearrow\nwarrow \omega_2$\hfil}
\line{\hskip 0.04in\hfil $\vert$ \hfil}
\line{\hskip 0.04in\hfil $\vert$ \hfil}

\line{\hskip 0.04in\hfil $\vert$ \hfil}

\line{\hskip 0.04in\hfil $\vert$ \hfil}

\line{\hskip 0.13in\hfil $\vert n$ \hfil}

\line{\hskip 0.04in\hfil $\vert$ \hfil}

\line{\hskip 1.50in {\hbox to 0.75in{\leftarrowfill}}$p_1${\hbox to 0.60in{\rightarrowfill}}$\vert$\kern -0.1em{\hbox to 0.75in{\leftarrowfill}}$p_2\sim${\hbox to 0.53in{\rightarrowfill}}\hskip 0.25in$\phi$\hfil}
\line{\hfil\hskip 0.10in$\omega_1 \swarrow$\kern -.6em\vbox{\hrule width .15in}$\theta$\kern -.6em\vbox{\hrule width 1.60in}{\vrule height6pt width1pt depth0pt}\kern -0.1em{\vrule height0pt width1pt depth6pt}\kern-.1em\vbox{\hrule width 1.5in}$\sim$\vbox{\hrule width 0.20in}\kern -.6em $\searrow \omega_2$\hfil}
\line{\hskip 1.05in$\leftarrow F_1 $\hskip 1.55in $\omega_3$\hskip 1.70in  $ F_2\rightarrow$\hfil}\par \bigskip

Since Einstein measures are to be associated with this diagram$,$ then this diagram should be obtained relative to infinitesimal light-clock counts and processes in the NSPPM.  The three locations $F_1,\ F_2,\ P$ are assumed$,$ at first$,$ to coincide. When this occurs$,$ the infinitesimal light-clock counts coincide. The positions $F_2, \ F_1$ recede from each other with velocity $\omega_3$. The positions $F_2, \ F_1$ recede from each other with velocity $\omega_3$. The object denoted by location $P$ recedes from the $F_1,\ F_2$ locations with uniform NSPPM velocities$,$ in standard form$,$ of $\omega_1,\ \omega_2$, respectively. Further, consider the special case where both are observing the pulse sent from $P$ at the exact some $P$-time. This produces the internal angle $\theta$ and exterior angle $\phi$ for this velocity triangle. 
The segments marked $p_1$ and $p_2$ are the projections of the velocity representations (not vectors) $F_1P$ and $F_2P$ onto the velocity representation $F_1F_2$. The $n$ is the usual normal for this projection. We note that $p_1 + p_2 = \omega_3.$  We apply hyperbolic 
trigonometry in accordance with [2]$,$ where we need to consider a particular $k$. We do this by scaling the velocities in terms of light units and let $k = c$. From [2, p. 143]\par

$$\cases{\tanh(p_1/c) = (\tanh (\omega_1/c)) \cos \theta&\cr
          \tanh (p_2/c) = - (\tanh (\omega_2/c))\cos \phi&\cr},\eqno (A4)$$
and also 
$$\sinh (n/c) = (\sinh (\omega_1/c))\sin \theta = (\sinh (\omega_2/c))\sin \phi. \eqno (A5)$$
Now$,$ eliminating $\theta$ from $(A4)$ and $(A5)$ yields [1, p. 146]
$$\cosh (\omega_1/c) = (\cosh (p_1/c))\cosh (n/c). \eqno (A6)$$
Combining (A4), (A5) and (A6) leads the hyperbolic cosine law [2, p. 167]. 
$$\cosh (\omega_1/c)=(\cosh (\omega_2/c))\cosh (\omega_3/c) + (\sinh(\omega_2/c))(\sinh(\omega_3/c))\cos \phi. \eqno (A7)$$
From $(A3),$ where each $v_i$ is the Einstein relative velocity$,$ we have that 
$$e^{\omega_i/c} = \left({{1 + v_i/c}\over{1 - v_i/c}}\right)^{(1/2)}, i = 1,2,3.\eqno (A3)'$$
From the basic hyperbolic definitions$,$ we obtain from $(A3)'$
$$\cases{\tanh (\omega_i/c) = v_i/c&\cr
          \cosh (\omega_i/c) = (1 - v_i^2/c^2)^{-1/2} = \beta_i&\cr
          \sinh (\omega_i/c) = \beta_iv_i/c&\cr}. \eqno (A8)$$
Our final hyperbolic requirement is to use 
$$\tanh (\omega_3/c) = \tanh(p_1/c + p_2/c)= {{\tanh (p_1/c) + \tanh (p_2/c)}\over{1 + (\tanh (p_1/c))\tanh (p_2/c)}}. \eqno (A9)$$\pars
Now into $(A9),$ substitute $(A4)$ and then substitute the first case from $(A8).$ One obtains
$$v_1\cos \theta = {{v_3 - v_2\cos \phi}\over{1 - \alpha}},\ \alpha = {{v_3v_2\cos \phi}\over{c^2}}. \eqno (A10)$$
Substituting into $(A7)$ the second and third cases from $(A8)$ yields
$$\beta_1=\beta_2\beta_3(1-\alpha), \ \beta_i = (1-v_i^2/c^2)^{-1/2}.\eqno (A11)$$
From equations $(A11), \ (A5)$ and the last case in $(A8)$ is obtained
$$v_1\sin \theta ={{v_2\sin\phi}\over{\beta_3(1-\alpha)}}. \eqno (A12)$$\par

For the specific physical behavior being displayed, the photons received from $P$ at $F_1$ and $F_2$ are ``reflected back'' at the  
NSPPM $P$-time $t^r.$ We then apply to this three point scenario our previous results. [Note: For comprehension, it may be necessary to apply certain relative velocity viewpoints such as from $F_1$ the point $P$ is receding from $F_1$ and $F_2$ is receding from $P$.  In this case, the NSPPM times when the photons are sent from $F_1$ and $F_2$ are related. Of course, as usual there is assumed to be no time delay between the receiving and the sending of a ``reflected''  photon.] In this case$,$ let $t^{(1)},\ r^{(1)},\ v_1$ be the Einstein measures at $F_1$ for this $P$-event$,$ and $t^{(2)},\ r^{(2)},\ v_2$ be the Einstein measures at $F_2$. Since $t^r = \beta_1^{-1}t^{(1)},\ t^r = \beta_2^{-1}t^{(2)}$ (p. 52), then  
$${{t^{(1)}}\over {\beta_1}} = {{t^{(2)}}\over {\beta_2}}\ {\rm and}\ r^{(1)}=v_1t^{(1)},\ r^{(2)}=v_2t^{(2)}.\eqno (A13)$$\par

Suppose that we have the four coordinates$,$ three rectangular$,$ for this $P$ event as measured from $F_1 = (x^{(1)}, y^{(1)}, z^{(1)}, t^{(1)})$ and from $F_2 = (x^{(2)}, y^{(2)}, z^{(2)}, t^{(2)})$ in a three point plane. It is important to recall that the $x,y,z$ are related to Einstein measures of distance. Further$,$ we take the $x$-axis as that of 
$F_1F_2$. The $v_3$ is the Einstein measure of the $F_2$ velocity as measured by an inf. light-clock at $F_1$. To correspond to the customary coordinate system employed [1, p. 32], this gives
$$\cases{x^{(1)}=v_1t^{(1)}\cos\theta,\ y^{(1)}=v_1t^{(1)}\sin\theta, \ z^{(1)} = 0&\cr
         x^{(2)}=-v_2t^{(2)}\cos\phi,\ y^{(2)}=v_2t^{(2)}\sin\phi, \ z^{(2)} = 0&\cr}.\eqno (A14)$$\pars

It follows from $(A10), \cdots, (A14)$ that 
$$t^{(1)} = \beta_3(t^{(2)} -v_3x^{(2)}/c^2),\ x^{(1)} = \beta_3(x^{(2)} -v_3t^{(2)}),\ y^{(1)} = y^{(2)},\ z^{(1)} = z^{(2)}. \eqno (A15)$$\pars

Hence, for this special case $\omega_1,\omega_2, \theta,\ \phi$ are eliminated and the Lorentz Transformations are established.  If $P \not= F_1,  P\not= F_2$, then the fact that $x^{(1)},\ x^{(2)}$ are not the measures for a physical ruler but are measures for a distance related to Einstein measures$,$ which are defined by the properties of the propagation of electromagnetic radiation and infinitesimal light-clock counts$,$ shows that the notion of actual \underbar{natural} world ``length'' contraction is false. For logical consistency$,$ Einstein measures as determined by the light-clock counts are necessary. This analysis is relative to a ``second'' pulse when light-clock counts are considered. The positions $F_1$ and $F_2$ continue to coincide during the first pulse light-clock count determinations. \par

Infinitesimal light-clock counts allow us to consider a real interval as an interval for ``time'' measure as well as to apply infinitesimal analysis. This is significant when the line-element method in Article 3 is applied to determine alterations in physical behavior. All of the coordinates being considered must be as they would be understood from the Einstein measure viewpoint. The interpretations must always be considered from this viewpoint as well. 
Finally, the model theoretic error of generalization is eliminated by predicting alterations in clock behavior rather than by the error of inappropriate generalization. \parm \parm
\centerline{\bf REFERENCES} \parm
\noindent {\bf Dingle, H.} {\it The Special Theory of Relativity,} Methuem \& Co., London, 1950.\pars
\noindent {\bf Herrmann, R. A.} ``The NSP-world and action-at-a-distance,'' In ``Instantaneous Action at a Distance in Modern Physics: `Pro' and `Contra,' ''
Edited by Chubykalo, A.,  N. V. Pope and R. Smirnov-Rueda,
(In CONTEMPORARY FUNDAMENTAL PHYSICS - V. V. Dvoeglazov (Ed.))
Nova Science Books and Journals, New York, 1999, pp. 223-235.  Also see http://arxiv.org/abs/math/9903082 pp. 118-119.\pars
\noindent {\bf Kulczycki, S.} {\it Non-Euclidean Geometry,} Pergamon Press, New York, 1961.

\end